\RequirePackage{mathtools}
\RequirePackage{commath, amsmath, amssymb, amsfonts}
\documentclass[10pt, a4paper]{article}
\usepackage[margin=1in]{geometry}
\usepackage{authblk}
\usepackage{float}

\newcommand{\onlinecite}[1]{\cite{#1}}

\usepackage{graphicx}
\usepackage{hyperref}
\hypersetup{
	colorlinks   = true, 
	urlcolor     = blue, 
	linkcolor    = blue, 
	citecolor   = blue 
}

\newcommand{\ie}{i.e.~}
\newcommand{\cf}{c.f.~}
\newcommand{\eg}{e.g.~}

\newcommand{\va}[1]{\mathbf{#1}}

\newcommand{\dd}[1]{\mathrm{d}{#1}}

\newcommand{\ket}[1]{|{#1}\rangle}

\newcommand{\mel}[3]{\langle{#1}|{#2}|{#3}\rangle}
\newcommand{\field}[1]{\boldsymbol{\mathcal{{#1}}}}
\newcommand{\vdot}{\boldsymbol\cdot}

\newcommand{\hin}{\va{u}_\mathrm{in}\vdot\hat{\va{P}}}
\newcommand{\hout}{\va{u}_\mathrm{out}\vdot\hat{\va{P}}}
\newcommand{\gnu}{\hat{G}_\nu}
\newcommand{\ele}[1]{\mathcal{#1}}
\newcommand{\irr}[1]{\mathrm{#1}}

\begin{document}

\title{A library of \textit{ab initio} Raman spectra for automated identification of 2D materials}

\author[1,2,3]{Alireza Taghizadeh\thanks{ata@nano.aau.dk}}
\author[3]{Ulrik Leffers}
\author[1,2]{Thomas G. Pedersen}
\author[3,4]{Kristian S. Thygesen}
\affil[1]{Department of Materials and Production, Aalborg University, 9220 Aalborg {\O}st, Denmark}
\affil[2]{Center for Nanostructured Graphene (CNG), 9220 Aalborg {\O}st, Denmark}
\affil[3]{Computational Atomic-scale Materials Design (CAMD), Department of Physics, Technical University of Denmark (DTU), 2800 Kgs. Lyngby, Denmark}
\affil[4]{Center for Nanostructured Graphene (CNG), Technical University of Denmark (DTU), 2800 Kgs. Lyngby, Denmark}
\makeatletter
\renewcommand*{\@fnsymbol}[1]{\ensuremath{\ifcase#1\or \dagger\or *\or \ddagger\or
    \mathsection\or \mathparagraph\or \|\or **\or \dagger\dagger
    \or \ddagger\ddagger \else\@ctrerr\fi}}
\makeatother
\maketitle

\begin{abstract}
Raman spectroscopy is frequently used to identify composition, structure and layer thickness of 2D materials.
Here, we describe an efficient first-principles workflow for calculating resonant first-order Raman spectra of solids within third-order perturbation theory employing a localized atomic orbital basis set. The method is used to obtain the Raman spectra of 733 different monolayers selected from the computational 2D materials database (C2DB).
We benchmark the computational scheme against available experimental data for 15 known monolayers. Furthermore, we propose an automatic procedure for identifying a material based on an input experimental Raman spectrum and illustrate it for the cases of MoS$_2$ (H-phase) and WTe$_2$ (T$^\prime$-phase). The Raman spectra of all materials at different excitation frequencies and polarization configurations are freely available from the C2DB. Our comprehensive and easily accessible library of \textit{ab initio} Raman spectra should be valuable for both theoreticians and experimentalists in the field of 2D materials. 	
\end{abstract}


\section*{Introduction \label{sec:Introduction}}

\noindent Following the discovery of graphene in 2004 \cite{Novoselov2004}, the field of two-dimensional (2D) materials has grown tremendously during the last decade. Today, more than 50 different monolayer compounds including metals \cite{Anasori2017,Xi2015}, semiconductors \cite{Liu2014, Mak2010, Wang2012}, insulators \cite{Ci2010}, ferromagnets \cite{Huang2017}, and superconductors \cite{Wang2017, Cao2018}, have been chemically grown or mechanically exfoliated from layered bulk crystals \cite{Haastrup2018}. The enormous interest in 2D materials has mainly been driven by their unique and easily tunable properties (as compared to 3D bulk crystals), which make them attractive for both fundamental research and technological applications in areas such as energy conversion/storage, (opto)-electronics, and photonics \cite{Bonaccorso2010, Wang2012, Gjerding2017}. Among the various experimental techniques used for characterizing 2D materials, Raman spectroscopy plays a pivotal role \cite{Ferrari2007} thanks to its simplicity, non-destructive nature, and high sensitivity towards key materials properties such as chemical composition, layer thickness (number of layers), inter-layer coupling, strain, crystal symmetries, sample quality, etc \cite{Ferrari2013, Neumann2015, Li2017}. 

\begin{figure}[t]
	\centering
	\includegraphics[width=0.9\textwidth]{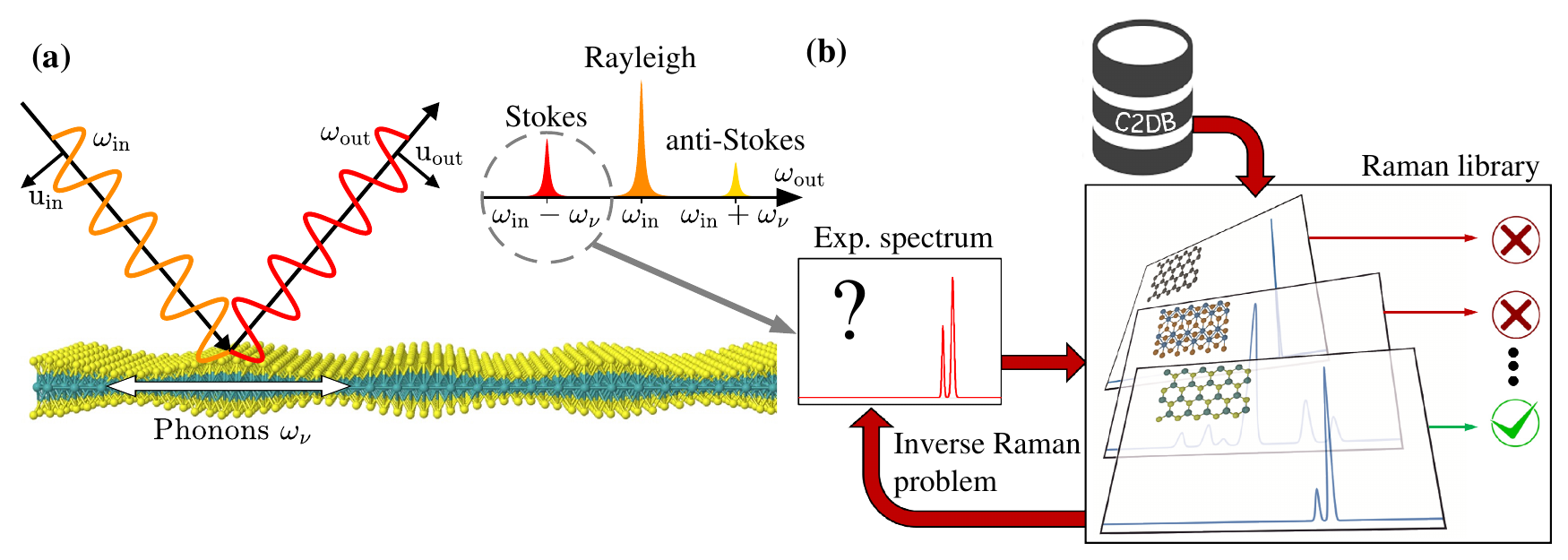}
	\caption[Schematic]{Schematic view of Raman scattering process and inverse Raman problem. (a) Raman scattering processes, in which incident photons of polarization $\va{u}_\mathrm{in}$ and frequency $\omega_\mathrm{in}$ are scattered into $\va{u}_\mathrm{out}$ and $\omega_\mathrm{out}$ under emission (or absorption) of a phonon with frequency $\omega_\nu$. Only zero momentum phonons contribute to first-order Raman processes but, for illustrative purposes, a finite momentum phonon is shown here. In a typical output spectrum, the Rayleigh (elastic), Stokes and anti-Stokes lines are observed. (b) Given an experimental spectrum, the Raman library based on the open Computational 2D Materials Database (C2DB) can be used to tackle the inverse Raman problem, \ie identifying the underlying material based on its Raman spectrum.} 
	\label{fig:Schematic}
\end{figure}

Raman spectroscopy is a versatile technique for probing the vibrational modes of molecules and crystals from inelastically scattered light, and is widely used for identifying materials through their unique vibrational fingerprints \cite{Long2002}. There are various types of Raman spectroscopies that differ in the number of photons or phonons involved in the scattering process \cite{Long2002}. Here we focus on the first-order Raman processes in which only a single phonon is involved. Typically, this is the dominant scattering process in defect-free samples (which are considered here). Note that Raman processes involving defect states or several phonons may also play important roles in some 2D crystals such as graphene \cite{Saito2011}. As shown schematically in Fig.~\ref{fig:Schematic}(a), the light scattered from a crystal appears in three distinct frequency bands: A strong resonance at the incident frequency $\omega_\mathrm{in}$ due to Rayleigh (elastic) scattering, and weaker resonances due to Raman (inelastic) scattering at $\omega_\mathrm{in}-\omega_\nu$ and $\omega_\mathrm{in}+\omega_\nu$ forming Stokes and anti-Stokes bands, respectively. Here, $\omega_\nu$ is the frequency of a (Raman active) vibrational mode of the crystal, \ie a phonon. Depending on the symmetry of the phonon modes and polarization of the electromagnetic fields, a phonon mode may be active or inactive in the Raman spectrum.

While semi-classical theories of Raman spectroscopy can provide some qualitative insight \cite{Long2002}, a full quantum mechanical treatment is necessary for a quantitatively accurate description. In particular, \textit{ab initio} techniques have been employed successfully to calculate Raman spectra of both molecules \cite{Wilson1980, Long2002} and solids \cite{Umari2003, Liang2019} typically showing good agreement with experimental spectra. The parameter-free nature of such computational schemes endow them with a high degree of predictive power, although their computational cost can be significant, thus, in practice limiting them to relatively simple, i.e. crystalline, materials. In the realm of 2D materials, \emph{ab initio} Raman studies have been limited to a handful of the most popular 2D crystals including graphene \cite{Saito2011}, hBN \cite{Cai2017}, WTe$_2$ \cite{Jiang2016}, SnS and SnSe \cite{Dewandre2019}, as well as MoS$_2$ and WS$_2$ \cite{Liang2014}. In view of the significant experimental efforts currently being devoted to the synthesis and application of future 2D materials and the important role of Raman spectroscopy as a main characterization tool, it is clear that the compilation of a comprehensive library of Raman spectra of 2D materials across different crystal structures and chemical compositions is a critical and timely endeavour. 

Recently, we have introduced the open Computational 2D Materials Database (C2DB) \cite{Haastrup2018}, which contains various calculated properties for several thousands 2D crystals using state of the art \textit{ab initio} methods. The properties currently provided in the C2DB include the relaxed crystal structures, thermodynamic phase diagrams (convex hull), electronic band structures and related quantities (effective masses, deformation potentials, etc.), elastic properties (stiffness tensors, phonon frequencies), and optical conductivity/absorbance spectra. We stress that the materials in the C2DB comprise both experimentally known as well as hypothetical but dynamically stable materials, i.e. materials that are predicted to be stable but may or may not be possible to synthesise in reality.

In this paper, we present an \textit{ab initio} high-throughput computation of the resonant first-order Raman spectra of more than 700 monolayers from the C2DB. The calculations are based on an efficient density functional theory (DFT) implementation of the first-order Raman process employing a localized atomic orbital (LCAO) basis set \cite{Larsen2009}. We describe the implementation and the automated workflow for computing the Raman spectra at three different excitation frequencies and nine polarization setups. All calculated Raman spectra are provided in Supplementary Figures~2-734, and can be found at the C2DB website \cite{C2DB}. In addition, the applied computational routines are freely available online through the website. Our numerical results are benchmarked against available experimental data for selected 2D crystals (15 different monolayers) such as MoS$_2$, MoSSe and MoSe$_2$. The calculated spectra show excellent agreement with experiments for the Raman peak positions and acceptable agreement for the relative peak intensities. Finally, we analyze the inverse problem of identifying a material based on an input (experimental) Raman spectrum as shown schematically in Fig.~\ref{fig:Schematic}(b). Using MoS$_2$ (H-phase) and WTe$_2$ (T$^\prime$-phase) as two examples, we find that a simple descriptor consisting of the first and second moments of the Raman spectrum combined with the Euclidean distance measure suffices to identify the correct material among the 700+ candidate materials in the database. In particular, this procedure can be used to differentiate clearly the distinct structural phases of MoS$_2$ and WTe$_2$. Incidentally, the library of calculated Raman spectra provides a useful dataset for training machine learning algorithms \cite{Butler2018, Schmidt2019}. As such, our work is not only a valuable reference for experimentalists and theoreticians working in the field of 2D materials, but also represents a step in the direction of autonomous (in-situ) characterization of materials.


\section*{Results}


\subsection*{Theory of Raman Scattering \label{sec:Theory}}

We first briefly review the theory of Raman scattering in the context of third-order perturbation theory.
As discussed above, accurate modeling of Raman processes requires a quantum mechanical treatment to obtain the electronic properties.
Regarding the electromagnetic field, it can be shown that a classical description of the field \cite{Albrecht1961, Lee1979} yields the same results as the full quantum mechanical theory that quantizes the photon field \cite{Long2002, Ganguly1967}. The most common approach to Raman calculations is the Kramers-Heisenberg-Dirac approach, \cite{Lee1979} in which the Raman tensor is obtained as a derivative of the electric polarizability with respect to the vibrational normal modes \cite{Albrecht1961, Lee1979, Umari2003, Liang2014}. Nonetheless, here we employ a more direct and much less explored approach based on time-dependent third-order perturbation theory to obtain the rate for coherent electronic processes involving creation/annihilation of two photons and one phonon. While the two approaches can be shown to be equivalent \cite{Long2002}, at least when local field effects can be ignored as is the case for 2D materials, the third-order perturbative approach can be readily extended to higher order Raman processes (\eg scattering on multiple phonons), and provides a more transparent physical picture of the Raman processes in terms of individual scattering events \cite{Yu2010}. Hence, our computational framework is prepared for future extensions to multi-phonon processes. Note that in terms of computational effort, the perturbative approach is comparable to the polarizability derivative method for typical crystals, for which the matrix element calculation dominates the computation time. In this case, both approaches scale as $N_\nu N_b^2$, where $N_\nu$ and $N_b$ denote the number of phonon modes and electronic bands, respectively.

To derive an expression for the Raman intensity, both electron-light and electron-phonon Hamiltonians are treated as perturbations (the exact forms of these Hamiltonians are given in the method section). A general time-dependent perturbation can be written as $\hat{H}'(t) \equiv \sum_{\omega_1} \hat{H}^\prime(\omega_1) \exp(-i\omega_1 t)$ ($\omega_1$ runs over positive and negative frequencies and can also be zero). Note that, in our study, there are three distinct frequency components in $\hat{H}'(t)$: input and output frequencies ($\omega_\mathrm{in}$ and $\omega_\mathrm{out}$) due to the electron-light interaction and zero frequency (\ie time-independent) for electron-phonon coupling. Within third-order perturbation theory, the transition rate $P_{i \rightarrow f}^{(3)}$ from an initial state $\ket{\Psi_i}$ to a final state $\ket{\Psi_f}$ due to the perturbative Hamiltonian $\hat{H}'(t)$, is given by \cite{Grosso2000}
\begin{align}
\label{eq:Pertubation}
&P_{i \rightarrow f}^{(3)} = \frac{2\pi}{\hbar} \abs{ \sum_{ab} \sum_{(\omega_1\omega_2\omega_3)} \frac{ \mel{\Psi_f}{\hat{H}^\prime(\omega_1)}{\Psi_a} \mel{\Psi_a}{\hat{H}^\prime(\omega_2)}{\Psi_b} \mel{\Psi_b}{\hat{H}^\prime(\omega_3)}{\Psi_i} }{(E_i-E_a+\hbar\omega_2+\hbar\omega_3)(E_i-E_b+\hbar\omega_3)} }^2 \delta(E_f-E_i-\hbar\omega) \, . 
\end{align}
Here, $a,b$ summations are performed over all eigenstates of the unperturbed system (here a set of electrons and phonons) and the sums over $\omega_n$ with $n=1,2,3$ are over all three involved frequencies in the perturbative Hamiltonian $\hat{H}'(t)$.
The notation $(\omega_1\omega_2\omega_3)$ indicates that, in performing the summation over $\omega_n$, the sum $\omega_1+\omega_2+\omega_3=\omega$ is to be held fixed.
In addition, $E_\alpha$ with $\alpha\in \left\lbrace i,f,a,b\right\rbrace $ denote the energies associated with $\ket{\Psi_\alpha}$ and the Dirac delta ensures energy conservation. The light field is written as $\field{F}(t) = \mathcal{F}_\mathrm{in}\va{u}_\mathrm{in} \exp(-i\omega_\mathrm{in} t)+\mathcal{F}_\mathrm{out}\va{u}_\mathrm{out} \exp(-i\omega_\mathrm{out} t)+$complex conjugate, where $\mathcal{F}_\mathrm{in/out}$ and $\omega_\mathrm{in/out}$ are the amplitudes and frequencies of the input/output electromagnetic fields, respectively, see Fig.~\ref{fig:Schematic}(a). In addition, $\va{u}_\mathrm{in/out}=\sum_\alpha u_\mathrm{in/out}^\alpha\va{e}_\alpha$ denote the corresponding polarization vectors, where $\va{e}_\alpha$ is the unit vector along the $\alpha$-direction with $\alpha\in \lbrace x,y,z \rbrace$.

We now specialize to the case where the initial and final states of the system are given by $\ket{\Psi_i}=\ket{0}\otimes\ket{n_\nu}$ and $\ket{\Psi_f}=\ket{0}\otimes\ket{n_\nu+1}$, respectively \cite{Ganguly1967} so that $E_f-E_i=\hbar\omega_\nu$. Here, $\ket{0}$ denotes the ground state of the electronic system and $\ket{n_\nu}$ is a state with $n_\nu$ phonons at frequency of $\omega_\nu$. In this case, the intensity of the Stokes Raman process for a phonon mode is proportional to $P_{i \rightarrow f}^{(3)}$, in which the transition rate involves a photon absorption, followed by an emission of a single phonon and photon. For this type of processes, $(\omega_1,\omega_2,\omega_3)$ are any permutation of $(\omega_\mathrm{in},-\omega_\mathrm{out},0)$, \eg $\omega_1=\omega_\mathrm{in}$, $\omega_2=-\omega_\mathrm{out}$, $\omega_3=0$ and five similar terms (all six terms contribute to the response at frequency of $\omega=\omega_\mathrm{in}-\omega_\mathrm{out}$). The total Raman intensity $I(\omega)$ is then obtained by summing over all possible final states, \ie  phonon modes $\nu$. 
Inserting the perturbative Hamiltonians [\cf Equations~(\ref{eq:Hamiltonian1})-(\ref{eq:Hamiltonian3}) in method section] in Equation~(\ref{eq:Pertubation}), the expression for the Stokes Raman intensity involving scattering events by only one phonon can be written
\begin{align}
\label{eq:RamanFinal}
&I(\omega) = I_0 \sum_\nu \frac{n_\nu+1}{\omega_\nu} \bigg|\sum_{\alpha\beta} u_\mathrm{in}^\alpha R_{\alpha\beta}^\nu  u_\mathrm{out}^\beta \bigg|^2 \delta(\omega-\omega_\nu) \, .
\end{align}
Here, $I_0$ is an unimportant constant (since Raman spectra are always reported normalized) that is proportional to the input intensity and depends on the input frequency, and $n_\nu$ is given by the Bose--Einstein distribution, \ie $n_\nu \equiv (\exp[\hbar\omega_{\nu}/k_BT]-1)^{-1}$ at temperature $T$. Due to momentum conservation, only phonons at the center of the Brillouin zone contribute to the one-phonon Raman processes \cite{Saito2011}. Furthermore, $R_{\alpha\beta}^\nu$ denotes the Raman tensor for phonon mode $\nu$, see method section. Equation~(\ref{eq:RamanFinal}) is used for computing the Raman spectra in this work for a given excitation frequency and polarization setup. It may be noted that one can derive a similar expression for the anti-Stokes Raman intensity by replacing $n_\nu+1$ by $n_\nu$ in Equation~(\ref{eq:RamanFinal}) and $\omega_\nu$ by $-\omega_\nu$ in Equation~(\ref{eq:RamanTensor}) in method section. Note, also, that the Raman shift $\hbar\omega$ is expressed in cm$^{-1}$  with 1 meV equivalent to 8.0655 cm$^{-1}$.


\subsection*{Computational Workflow}

\begin{figure}[t!]
	\centering
	\includegraphics[width=0.85\textwidth]{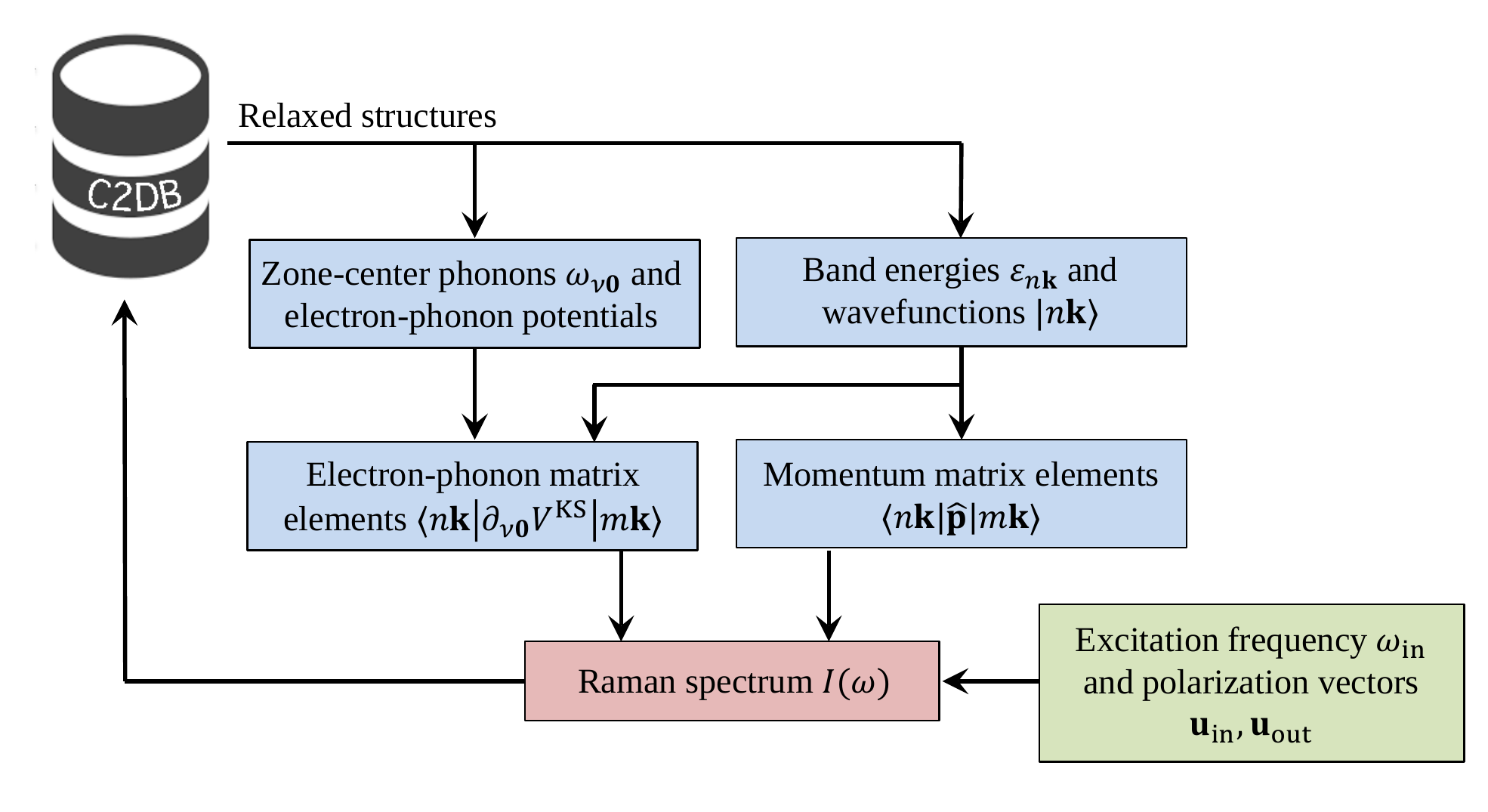}
	\caption[Diagram of the Raman calculation code.]{Computational workflow. The diagram illustrates the steps necessary to calculate the Raman tensor of a material. }
	\label{fig:Workflow}
\end{figure} 

An overview of the automated workflow for computing the Raman tensor of the materials in the C2DB is shown in Fig.~\ref{fig:Workflow}. First, the relaxed structures are extracted from the database. Next, the electronic band energies and wave functions are obtained from a DFT calculation. In parallel, a zone-center phonon calculation is performed to obtain the optical vibrational modes. From the obtained electronic states and phonon modes, the momentum and electron-phonon matrix elements are evaluated and stored. In the final step, for a given excitation frequency and input/output polarization vectors, the Raman spectrum is calculated using Equation~(\ref{eq:RamanFinal}). The key feature of the approach outlined here is that the calculation process can be automatized, allowing one to perform thousands of calculations in parallel without human intervention. 

\begin{figure}[t]
	\centering
	\includegraphics[width=0.9\textwidth]{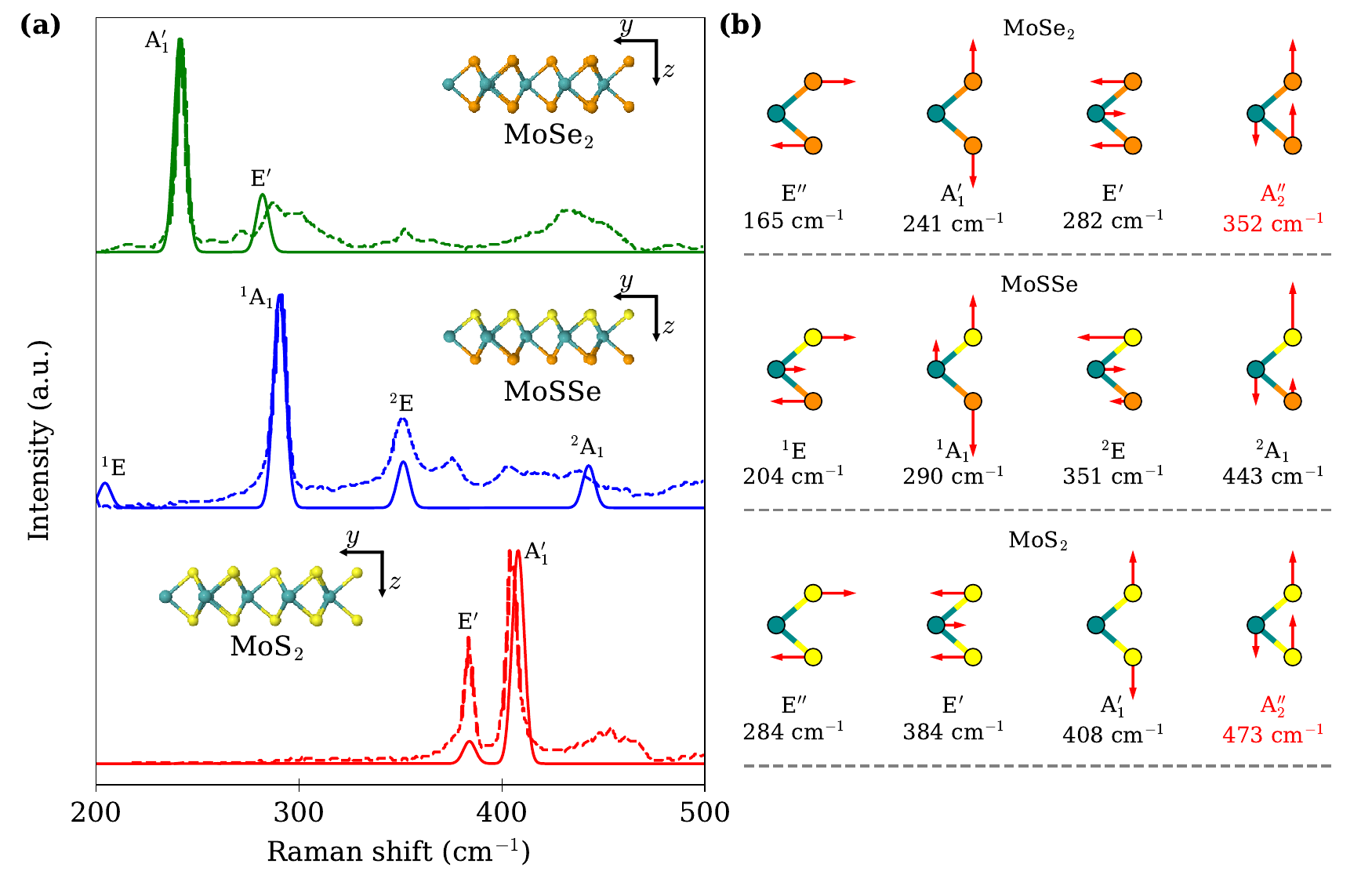}
	\caption[Raman spectra of the MoSe$_2$, MoSSe and MoS$_2$.]{Evolution of Raman spectra from MoSe$_2$ over MoSSe to MoS$_2$. (a) Comparison of the computed Raman spectra (solid) with the experimental results in Ref.~\cite{Zhang2017} (dashed) for MoSe$_2$ (top), MoSSe (middle) and MoS$_2$ (bottom). The excitation wavelength is 532 nm, and both input and output electromagnetic fields are polarized along the $y$-direction. (b) Optical phonon modes for MoSe$_2$ (top), MoSSe (middle) and MoS$_2$ (bottom) labeled by the irreducible representations of the respective point groups. Note that A$_2^{\prime\prime}$ modes (shown in red) are Raman inactive. }
	\label{fig:Janus}
\end{figure}

\begin{figure}[t]
	\centering
	\includegraphics[width=0.9\textwidth]{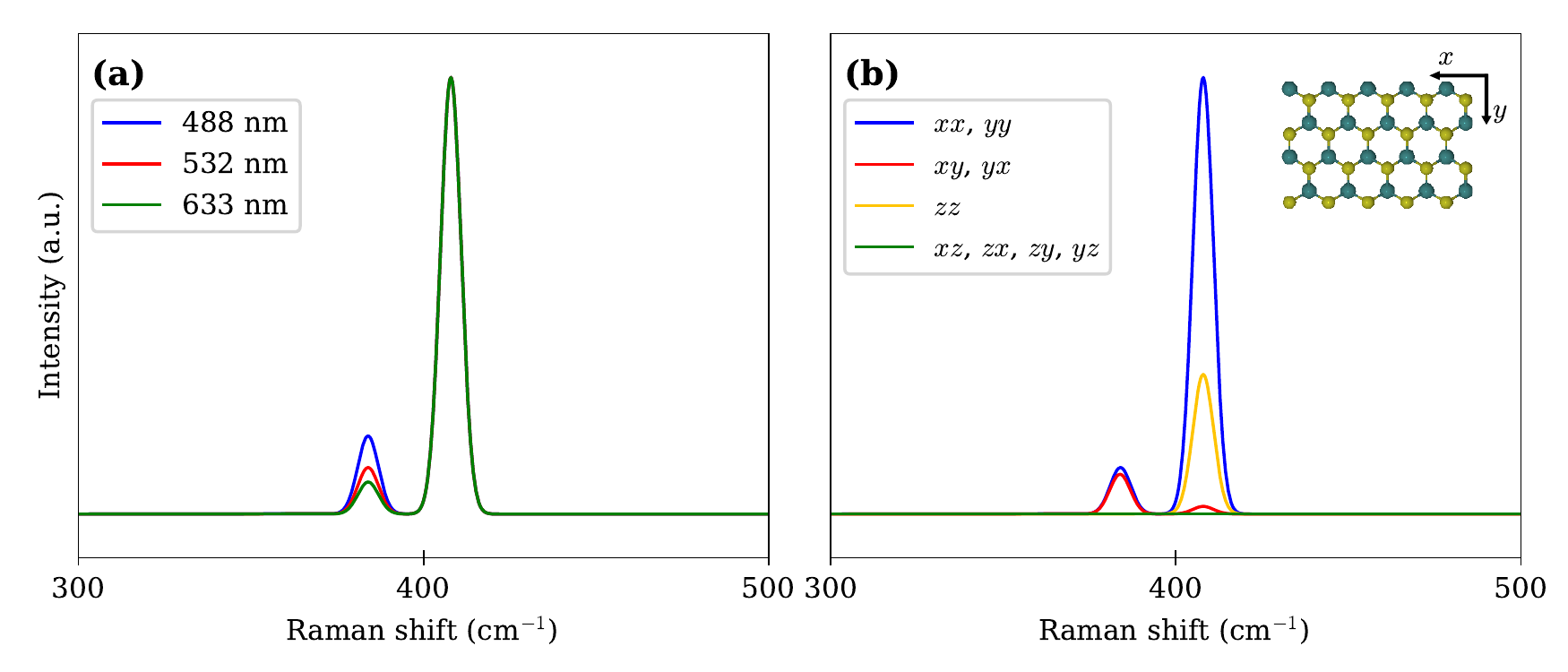}
	\caption[Raman spectra of MoS$_2$]{Polarization and frequency dependent Raman spectra. (a) Raman spectra of MoS$_2$ evaluated at three different excitation wavelengths, blue (488 nm), green (532 nm), and red (633 nm) for the $xx$-polarization setup. (b) Polarized Raman spectra of MoS$_2$ for various input and output polarization directions at 532 nm excitation wavelength. The inset shows a top view of the crystal structure. }
	\label{fig:Polarized}
\end{figure}

For simplicity, we have restricted the study to non-magnetic materials, but our routines can be readily extended to include magnetic materials. The Raman spectra presented in this paper are computed for in-plane polarization, where the incoming and outgoing photons are polarized along the $x$- or $y$-directions, \ie $\va{u}_\mathrm{in/out}$ are either $[1, 0, 0]$ or $[0, 1, 0]$. The four possible combinations are referred to as $xx$, $xy$, $yx$ and $yy$ polarization setups.


\subsection*{Raman Spectra and Comparison with Experiments}

Figure~\ref{fig:Janus} compares the calculated Raman spectrum of three different monolayer transition metal dichalcogenides (TMDs), namely MoSe$_2$, MoSSe and MoS$_2$, with the experimental data extracted from Ref.~\onlinecite{Zhang2017}. For all three monolayers, a good agreement is observed both for the peak positions and relative amplitudes of the main peaks. Additional peaks in the experimental spectra presumably originate from the substrate or defects in the samples. The differences between the Raman spectra of the three materials provide valuable information about the crystal structure. Symmetry and Raman activity of phonon modes are determined by the irreducible point group representations. MoS$_2$ and MoSe$_2$ are members of point group $D_{3h}$, whereas MoSSe lacking a horizontal mirror plane $\sigma_h$ belongs to the point group $C_{3v}$. In Mulliken notation, the irreducible representation of MoS$_2$ and MoSe$_2$ is $2\irr{A}_2''+\irr{A}_1'+2\irr{E}'+\irr{E}''$, whereas for MoSSe the lowered symmetry leads to $3\irr{A}_1+3\irr{E}$. For MoS$_2$ and MoSe$_2$, one member of both $\irr{A}_2''$ and $\irr{E}'$ is an acoustic mode, and the other $\irr{A}_2''$ mode is Raman inactive. For MoSSe, $\irr{A}_1$ and $\irr{E}$ each contain an acoustic mode, and all other modes are Raman active. The relevant modes are shown schematically in Fig.~\ref{fig:Janus}(b). In general, a Raman active mode will only appear in certain polarization configurations. The tensorial Raman selection rules follow from the irreducible point group representations \cite{Loudon1964, Ribeiro2014} as shown for point groups $D_{3h}$ and $C_{3v}$ in Supplementary Note 1.

\begin{figure}[p]
	\centering
	\includegraphics[width=0.9\textwidth,height=\textheight,keepaspectratio]{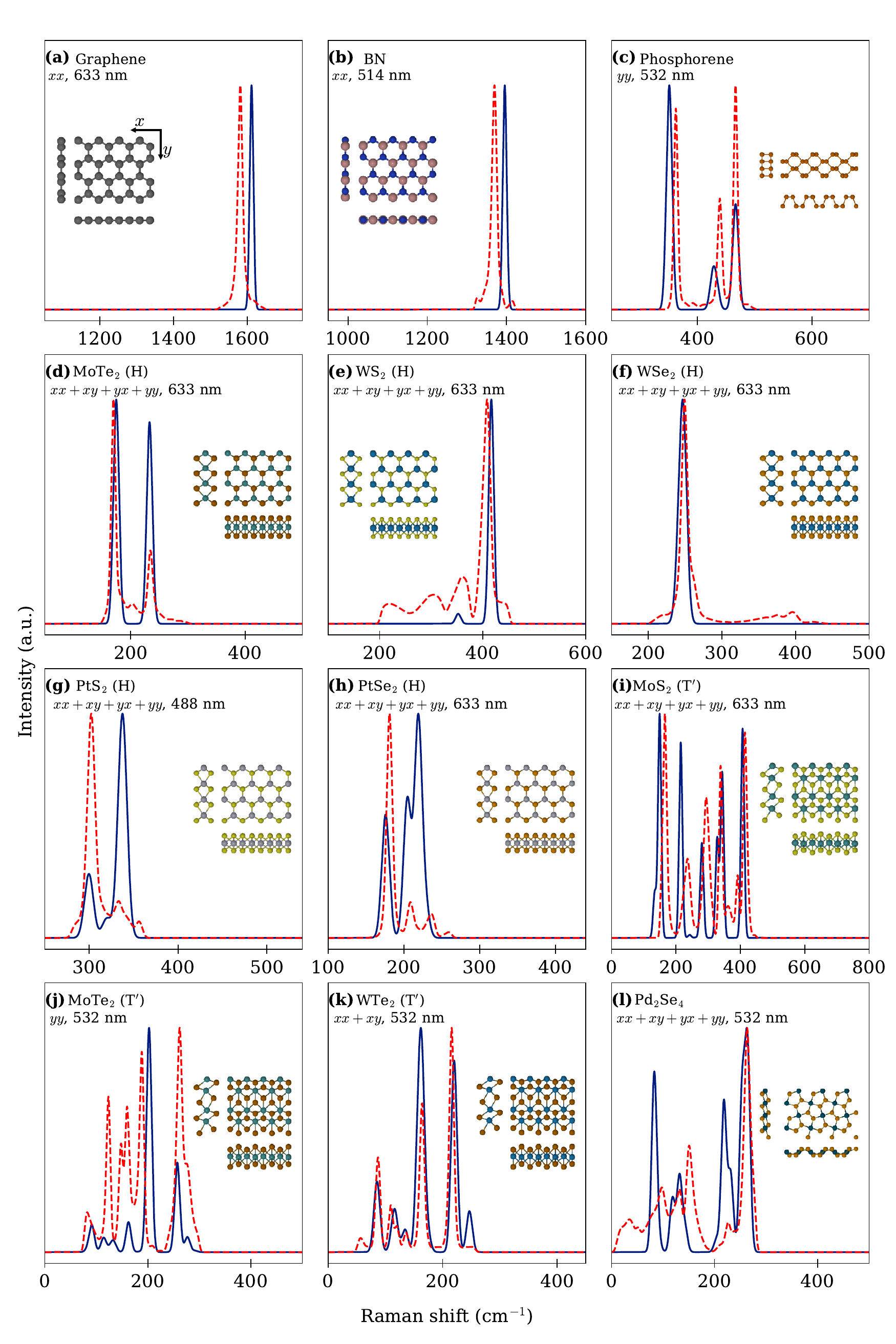}
	\caption[Compare with experiments.]{Raman spectra of 12 monolayers. Comparison of computed Raman spectra (solid lines) with available experimental results (dashed lines). The experimental data are extracted from Refs.~\onlinecite{Wu2018}, \onlinecite{Cai2017}, \onlinecite{Xia2014}, \onlinecite{Ruppert2014}, \onlinecite{Liu2018}, \onlinecite{Yang2017}, \onlinecite{Zhao2016b}, \onlinecite{Yan2017}, \onlinecite{Gupta2014}, \onlinecite{Chen2017}, \onlinecite{Cao2017}, and \onlinecite{Yu2020} for (a) to (l), respectively. The temperature is set to 300 K (room temperature) and excitation wavelength is specified in each case, see the main text. The crystal structures are shown in the insets including top view and cross sectional views. For all crystal structures the $x$- and $y$-directions are along the horizontal and vertical directions, respectively, as shown for graphene.}
	\label{fig:ExpCompare}
\end{figure}

Next, we focus on the case of MoS$_2$, and investigate the dependency of the Raman spectrum on the excitation frequency and polarization, see Figs.~\ref{fig:Polarized}(a) and \ref{fig:Polarized}(b), respectively. In Fig.~\ref{fig:Polarized}(a), the Raman spectra are computed for three commonly used  wavelengths of blue, green and red laser sources. In this case both in- and outgoing polarization vectors are along the $y$-direction (or $x$-direction). While the relative strength of the first Raman active peak in the spectrum is enhanced slightly for shorter wavelengths, the shape of the spectrum does not change significantly. Note that, in reality, the relative amplitudes of the $\irr{E}'$ and $\irr{A}_1'$ modes may change considerably if the excitation frequency coincides with an exciton resonance \cite{Carvalho2015, Scheuschner2015}. This is because excitons can selectively enhance specific Raman modes due to their symmetry \cite{Lee2015, Delcorro2016}. Although this effect is not captured properly in our independent-electron model, it is in principle straightforward to include by using the many-body eigenstates obtained by diagonalizing the Bethe-Salpeter equation (BSE) when evaluating the matrix elements in Equation~(\ref{eq:Pertubation}) \cite{Gillet2013, Delcorro2016, Wang2018, Taghizadeh2019, Taghizadeh2019b}. Moreover, the absolute magnitudes of the Raman peaks can vary substantially by changing the excitation wavelength due to the possible resonance with electronic states (resonance Raman spectroscopy) \cite{Lee2015}. Nonetheless, the overall magnitude of the Raman spectra is usually of little practical importance compared to the spectral positions and spectra are typically normalized as done here. Changing the polarization of electromagnetic fields not only influences the relative amplitudes of Raman peaks, but may switch certain modes on and off as shown in Fig.~\ref{fig:Polarized}(b). For instance, the MoS$_2$ $\irr{E}'$ mode becomes completely inactive for the perpendicular polarization setup ($zz$) due to symmetry \cite{Liang2014}.  
This is easily confirmed using Supplementary Equation~(1) of the Supplementary Note 1 predicting an inactive $\irr{E}'$ mode for $zz$-polarization. Note that, although the $\irr{E}''$ mode is Raman active for $xz$-, $yz$-, $zx$- and $zy$-polarizations, the intensity is too small to be observed in Fig.~\ref{fig:Polarized}(b).

We have assessed the quality of the Raman library for a wide range of material compositions and crystal structures. Figure~\ref{fig:ExpCompare} compares experimental and calculated Raman spectra for 12 monolayers including graphene, hBN, several conventional TMDs in the H- or T$^\prime$-phase as well as anisotropic crystals such as phosphorene and Pd$_2$Se$_4$. In general, the number of Raman active modes increases with the number of atoms in the unit cell, as expected. For instance, there are more than 8 peaks in the Raman spectrum of Pd$_2$Se$_4$. Furthermore, as a rule of thumb, Raman modes of materials containing heavier atoms are at lower frequencies and vice versa, \eg the Raman peaks for graphene and hBN appear at frequencies above 1000 cm$^{-1}$. The experimental data are obtained under various experimental conditions such as different excitation wavelengths and polarizations or diverse sample substrates. Note that if polarized Raman spectra were not available (or in the case of unspecified polarization), an average of all four in-plane polarization settings, \ie $xx$, $xy$, $yx$ and $yy$, has been used for generating the theoretical spectra. In general, there is quite good agreement between our calculations and experimental results in all cases, particularly, for the peak positions. The deviations can be attributed to various factors such as substrate and excitonic effects, which are not captured in our calculations, as well as the quality of the experimental samples and other experimental uncertainties, all of which can influence the spectra considerably.


\subsection*{Identifying Materials from Their Raman Spectra}

At this point, we turn to a critical test of the \textit{ab initio} Raman library: given an experimental Raman spectrum, is it possible to identify the underlying material by comparing the experimental spectrum to a library of calculated spectra? The answer to this question will depend on several factors including: (1) The quality of the experimental spectrum. (2) The quality of the calculated spectra, \ie the ability of theory to reproduce a (high quality) experimental spectrum for a given material. (3) The size/density of the calculated Raman spectrum database. Obviously, a more densely populated database increases the chances that the experimental sample is, in fact, contained in the database. But, at the same time, this increases the risk of obtaining a false positive, \ie matching the experimental spectrum by a calculated spectrum of a different material.  

Putting the above idea into practice requires a quantitative measure for comparing Raman spectra. In the present work, we use the two lowest moments to fingerprint the Raman spectrum. In general, the N$^{th}$ Raman moment of the spectrum is given by
\begin{equation}
\label{eq:Measures}
\langle\omega^N\rangle \equiv \int_{0}^\infty I(\omega) \omega^N \dd{\omega} = \sum_{\nu} I_\nu \omega_\nu^N \, ,
\end{equation}
where, $I_\nu$ denotes the amplitude of mode $\nu$, \ie $I_\nu=I_0(n_\nu+1)|\sum_{\alpha\beta} u_\mathrm{in}^\alpha R_{\alpha\beta}^\nu u_\mathrm{out}^\beta|^2/\omega_\nu$. Note that, for these calculations, we normalize the Raman spectrum such that its zeroth moment becomes one, \ie $\int_0^\infty I(\omega) \dd{\omega}=\sum_\nu I_\nu = 1$. Therefore, the first Raman moment corresponds to the mean value of the spectrum. Rather than using the second moment, we use the standard deviation of the spectrum as the selected measure, given by 
\begin{equation}
\label{eq:Measures2}
\delta \omega = \sqrt{\langle\omega^2\rangle-\langle\omega\rangle^2} \, .
\end{equation}

\begin{figure}[t]
	\centering
	\includegraphics[width=0.85\textwidth]{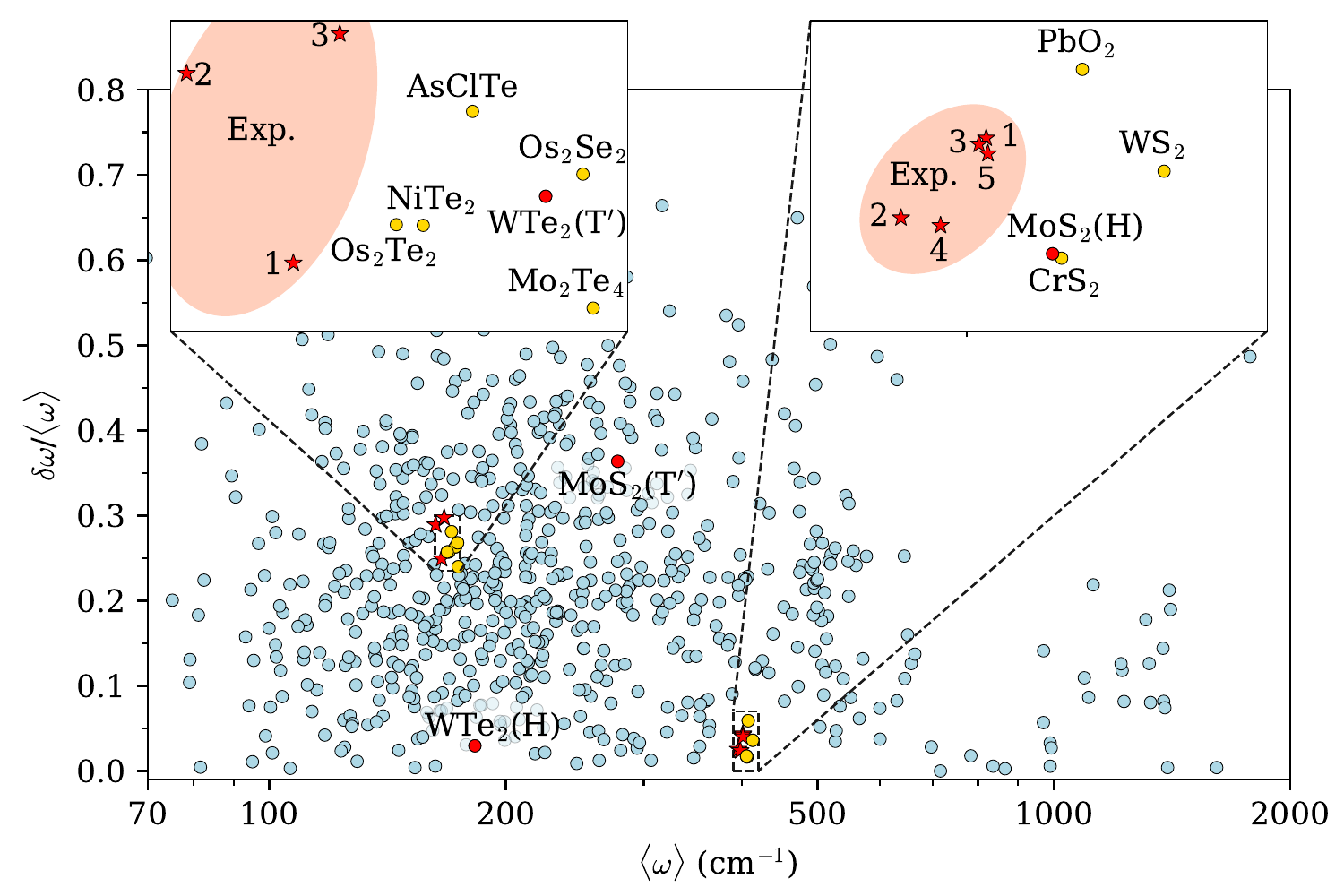}
	\caption[Scatter plot of calculated Raman spectra]{Calculated Raman moments. Scatter plot of the first Raman moment and normalized standard deviation for 733 calculated spectra at excitation wavelength of 532 nm and $xx$ polarization setup (circles). For comparison, several independent experimental spectra for monolayer MoS$_2$ (in H-phase) and WTe$_2$ (in T$^\prime$-phase) are also shown (stars). We highlight the points corresponding to MoS$_2$ and WTe$_2$ in red (H-phase, T$^\prime$-phase, and experiments). The insets are zooms of the vicinity of the experimental data for MoS$_2$ and WTe$_2$. For MoS$_2$, 1 to 5 correspond to the experimental spectra obtained from Refs.~\onlinecite{Zhang2017}, \onlinecite{Li2012}, \onlinecite{Lanzillo2013}, \onlinecite{Tongay2013}, and \onlinecite{Lee2015}, respectively, whereas 1 to 3 for WTe$_2$ are adopted from Refs.~\onlinecite{Chen2017}, \onlinecite{Zhou2016}, and \onlinecite{Cao2017}, respectively.}
	\label{fig:Scatter}
\end{figure}

Figure~\ref{fig:Scatter} shows a scatter plot of $\langle\omega\rangle$ and $\delta \omega/\langle\omega\rangle$ for the 733 monolayers at an excitation wavelength of 532 nm and $xx$ polarization setup obtained at the room temperature. In this plot, crystals composed of lighter elements appear further to the right because their optical phonons generally have higher energies. Furthermore, crystals with fewer atoms in the unit cell and/or higher degree of symmetry, appear in the bottom of the plot because they have fewer (non-degenerate) phonons and thus fewer peaks in their Raman spectrum resulting in a reduced frequency spread. In particular, $\delta \omega$ vanishes for materials with only a single Raman peak such as graphene and hBN. 

To test the feasibility of inverse Raman mapping, we evaluate the lowest Raman moment fingerprint for five experimental Raman spectra of MoS$_2$ (H-phase) and three spectra of WTe$_2$ (T$^\prime$-phase) obtained from independent studies, see stars in Fig.~\ref{fig:Scatter}. Similar analyses have been performed for the eleven additional crystals found in Fig.~\ref{fig:ExpCompare}, and is provided in the Supplementary Note 2. Firstly, note that the fingerprint of MoS$_2$ in the T$^\prime$-phase (WTe$_2$ in H-phase) is located relatively far from the H-phase (T$^\prime$-phase) fingerprint in the plot, which suggests that the lowest Raman moments are indeed able to distinguish different structural phases of the same material. The insets highlight the regions surrounding the experimental data. The variation in the experimental fingerprints is due to small differences in the Raman spectra, originating from the variations in sample quality, substrate effects, measurement techniques/conditions, etc. Consequently, the precise peak positions and, in particular, their amplitudes can vary from one experiment to another. Clearly, the fingerprints of the calculated spectra for both MoS$_2$ and WTe$_2$ lie close to the experimental data. In a few cases, such as Pd$_2$Se$_4$, the experimental fingerprints lie further from the theoretical predictions, as illustrated in the Supplementary Figure~1. This may partly be due to insufficient sample quality for these less-explored 2D crystals. In fact, the deviation between theory and experiments is comparable to the variation between the different experiments. Importantly, only a few other materials show a similar agreement with the experimental data. This suggests that fingerprints including higher order moments could single out the correct material with even higher precision. For instance, the skewness (based on the third Raman moment) can be used to distinguish MoS$_2$ from CrS$_2$. By manual inspection of the Raman spectra, one readily confirms that the calculated spectra of MoS$_2$ and WTe$_2$ are in fact the best match to the experimental spectra, \eg other candidates have Raman peaks that are not observed in the experimental spectra or the relative amplitudes of the peaks are completely different from the experimental data. Nonetheless, the procedure of manual inspection can be replaced by a more rigorous and unbiased  approach as discussed below.

\begin{figure}[t]
	\centering
	\includegraphics[width=0.85\textwidth]{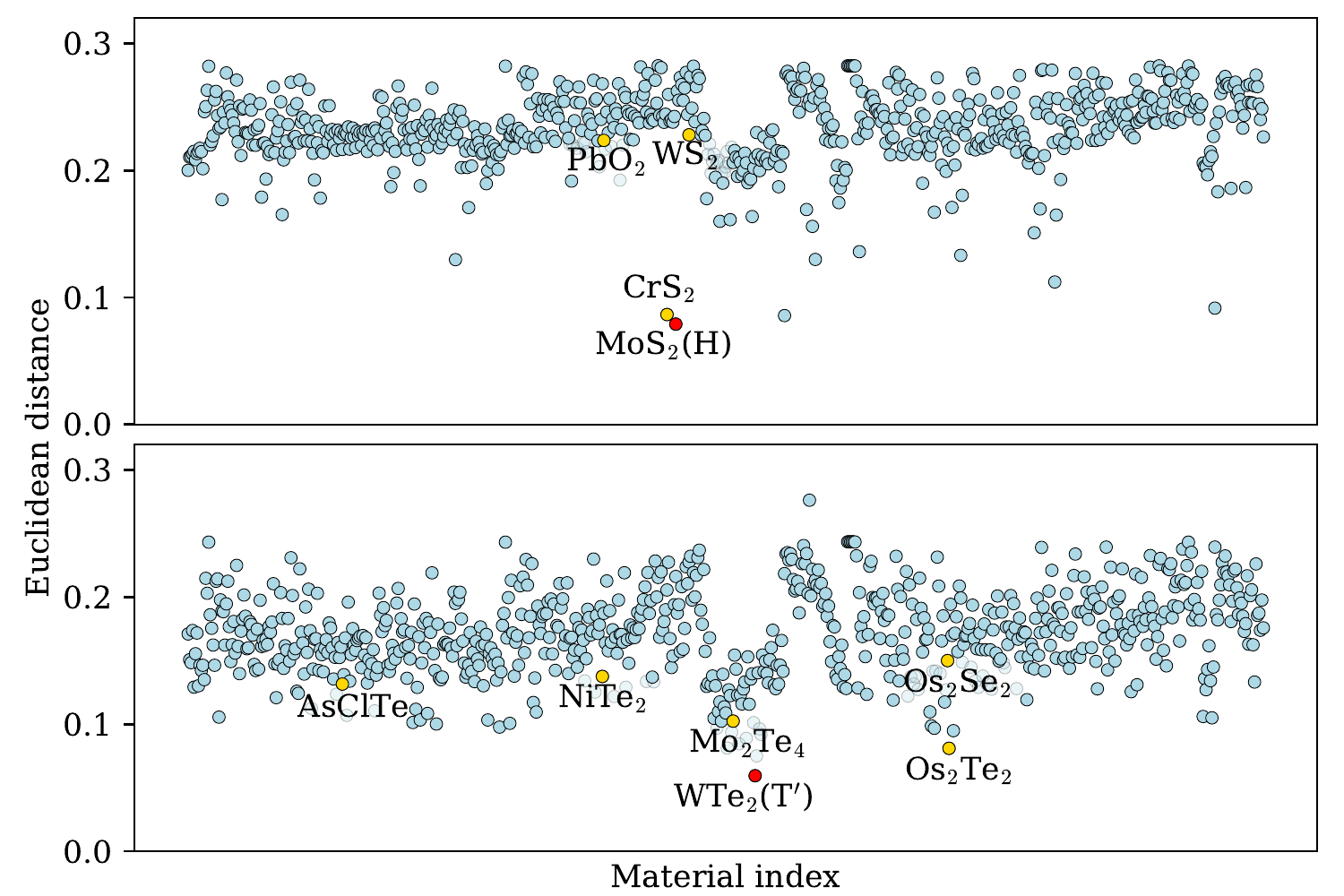}
	\caption[Distances of calculated Raman spectra]{Euclidean distances between Raman spectra. Distances (see main text for details) are calculated between theoretical Raman spectra and the experimental data of Refs.~\onlinecite{Tongay2013} and \onlinecite{Cao2017} for MoS$_2$ (top) and WTe$_2$ (bottom). For comparison purposes, we highlight the points corresponding to materials in the insets of Fig.~\ref{fig:Scatter} by yellow. }
	\label{fig:Distances}
\end{figure}

To compare the experimental and calculated Raman spectra quantitatively, we focus on the experimental data of Refs.~\onlinecite{Tongay2013} and \onlinecite{Cao2017} for MoS$_2$ and WTe$_2$, respectively. The experimental spectra for MoS$_2$ are obtained without any polarizer at 77 K at an excitation wavelength of 488 nm. For WTe$_2$ in Ref.~\onlinecite{Cao2017}, the experiment is performed  at room temperature using a 532 nm laser linearly polarized in-plane. To account for the unspecified polarization, we take the average of Raman spectra for the $xx$ and $xy$ polarization setups in the case of WTe$_2$, while for MoS$_2$ the average of all Raman spectra for transverse components ($xx$, $xy$, $yx$ and $yy$) is used as the theoretical spectrum. For quantitative comparison with the experimental data, one can use Euclidean distances between the experimental and theoretical spectra as a measure. For two Raman spectra $I_1(\omega$) and $I_2(\omega)$, the Euclidean distance (or $L^2$-norm) $||I_1-I_2||$ is defined as
\begin{equation}
\label{eq:Distances}
||I_1-I_2|| \equiv \Big( \int_0^\infty \abs{I_1(\omega)-I_2(\omega)}^2 \dd{\omega} \Big)^{1/2} \, .
\end{equation}
Note that the spectra are normalized such that the total area is unity. Figure~\ref{fig:Distances} shows the computed Euclidean distances from the calculated Raman spectra to the experimental data for both MoS$_2$ and WTe$_2$. We highlight the points corresponding to the materials in the insets of Fig.~\ref{fig:Scatter}. In both cases, identifying the smallest Euclidean distance confirms that the Raman spectra closest to the experimental data are indeed the calculated spectra of MoS$_2$ and WTe$_2$. This shows that the quality and accuracy of, respectively, the experimental and computed 2D materials Raman spectra, is sufficient for automatic structure identification.

\section*{Discussion \label{sec:Conclusions}}

We have introduced a comprehensive library of \textit{ab initio} computed Raman spectra for more than 700 2D materials spanning a variety of chemical compositions and crystal structures. The 2D materials comprise both experimentally known and hypothetical compounds, all dynamically stable and with low formation energies. Using an efficient first-principles implementation of third-order perturbation theory, the full  resonant first-order Raman tensor was calculated including all nine possible combinations for polarization vectors of the input/output photons and three commonly used excitation wavelengths. All spectra are freely available as part of the C2DB and should comprise a valuable reference for both theoreticians and experimentalists in the field. The reliability of the computational approach was demonstrated by comparison with experimental spectra for 15 monolayers such as graphene, hBN, phosphorene and several TMDs in the H-, T- and T$^\prime$-phases.    

We carefully tested the feasibility of inverse Raman mapping, \ie to what extent the library of computed Raman spectra can be used to identify the composition and crystal structure of an unknown material from its Raman spectrum. For the specific cases of MoS$_2$ in H-phase and WTe$_2$ in T$^\prime$-phase, we showed that a simple fingerprint based on the lowest moments of the Raman spectrum is sufficient to identify the materials from their experimental Raman spectrum. This represents a significant step in the direction of autonomous identification/characterization of materials. In addition, apart from being a useful reference for 2D materials research, the Raman library can be used to train machine learning algorithms to predict Raman spectra directly from the atomic structure similarly to recent work on prediction of linear optical spectra for molecules \cite{Ghosh2019}. This is of particular importance in the currently attractive trend of employing machine learning algorithms in materials science \cite{Butler2018, Schmidt2019}.

In the present work, we have focused on Raman processes involving only a single phonon, \ie first-order Raman processes, since these are typically the dominant contributions to the Raman spectrum. Nonetheless, the presented methodology can be readily extended to include two-phonon scattering processes, although the computational cost will be significantly increased. Excitonic effects in the Raman spectrum have been neglected since most experimental Raman spectra are recorded off-resonance where excitons play a minor role. The inclusion of excitonic effects can be achieved within the presented methodology by employing the many-body eigenstates obtained from the BSE \cite{Gillet2013, Cudazzo2016, Taghizadeh2018} instead of Slater determinantal electron-hole excitations. However, this will mainly affect the amplitude of the Raman peaks which is of secondary importance in practice. We only compute the Raman spectra of monolayers in the present work, but the library can be extended to multi-layer structures. For some 2D materials such as graphene or MoS$_2$ this can be done by employing existing exchange-correlation functionals capable of accurate modeling of van der Waals forces. But for other 2D systems such as phosphorene, further development of exchange-correlation functionals is required to describe the complex inter-layer couplings, particularly for low-frequency Raman modes \cite{Liang2017}. The symmetry of phonons modes have previously been investigated for graphene \cite{Malard2009}, the TMD family \cite{Ribeiro2014} and phosphorene \cite{Ribeiro2015} using group theory analysis. Based on the Raman library, such analysis could be performed for a much wider range of materials in future work. Finally, the current work has been restricted to non-magnetic materials, and the \textit{ab initio} Raman response of magnetic materials is an interesting future research field.


\section*{Methods \label{sec:Methods}}


\subsection*{Theory}
In the independent-particle approximation, the Hamiltonian of a system of electrons interacting with phonons and electromagnetic fields takes the form $\hat{H}=\hat{H}_0+\hat{H}_\mathrm{e\gamma}+\hat{H}_\mathrm{e\nu}$, where $\hat{H}_0$ is the unperturbed Hamiltonian of the electrons (e) and phonons ($\nu$), $\hat{H}_\mathrm{e\gamma}$ describes the electron-light interaction (here written in the velocity or minimal coupling gauge \cite{Taghizadeh2017, Reichardt2019}), and $\hat{H}_\mathrm{e\nu}$ describes the electron-phonon coupling. In second quantization, they are given by \cite{Giustino2017}
\begin{align}
\label{eq:Hamiltonian1}
&\hat{H}_0 \equiv \sum_{n\va{k}} \varepsilon_{n\va{k}} \hat{c}_{n\va{k}}^\dagger \hat{c}_{n\va{k}}+ \sum_{\nu\va{q}} \hbar \omega_{\nu\va{q}} (\hat{a}_{\nu\va{q}}^\dagger \hat{a}_{\nu\va{q}} +\frac{1}{2}) \, , \\
\label{eq:Hamiltonian2}
&\hat{H}_\mathrm{e\gamma}(t) = \dfrac{e}{m} \field{A}(t)\vdot \sum_{nm\va{k}} \va{p}_{nm\va{k}} \hat{c}_{n\va{k}}^\dagger \hat{c}_{m\va{k}}  \, , \\
\label{eq:Hamiltonian3}
&\hat{H}_\mathrm{e\nu} = \sum_{\substack{nm\nu \\ \va{k}\va{q}} } \sqrt{\frac{\hbar}{\omega_{\nu\va{q}}}} g_{nm\va{k}}^{\nu\va{q}} \hat{c}_{n\va{k}}^\dagger \hat{c}_{m\va{k}} (\hat{a}_{\nu\va{q}}+\hat{a}_{\nu(-\va{q})}^\dagger) \, .   
\end{align}
Here, $\hat{c}^\dagger/\hat{c}$ and $\hat{a}^\dagger/\hat{a}$ are the creation/annihilation operators of electrons and phonons, respectively, $\field{A}$ denotes the vector potential ($\field{F}=-\partial \field{A}/\partial t$), $\varepsilon_{n\va{k}}$ is the energy of the single-particle electronic state $\ket{n\va{k}}$, and $\hbar\omega_{\nu\va{q}}$ denotes the phonon energy of normal mode $\nu$ and wavevector $\va{q}$. Furthermore, $\va{p}_{nm\va{k}}=\mel{n\va{k}}{\hat{\va{p}}}{m\va{k}}$ and $g_{nm\va{k}}^{\nu\va{q}}=\mel{n\va{k}+\va{q}}{\partial_{\nu\va{q}} V^\mathrm{KS}}{m\va{k}}$ are the momentum and electron-phonon matrix elements (to the first order in the atomic displacements \cite{Giustino2017}), respectively, with the Kohn-Sham potential $V^\mathrm{KS}$. The summation over $\va{k}$ implies an integral over the first Brillouin zone, \ie $(2\pi)^D \sum_\va{k} \rightarrow V^D \int_\mathrm{BZ} \mathrm{d}^D{\va{k}}$ where $V$ is the $D$-dimensional volume ($D=2$ for 2D systems). Note that the $\field{A}^2$ term does not contribute to the linear Raman response and, hence, is absent here. Moreover, we neglect the Coulomb interaction between electrons and holes, \ie excitonic effects. If the Raman spectroscopy is performed with an excitation frequency that matches the exciton energy \cite{Carvalho2015}, the electron-hole interactions should be included, ideally within the GW and BSE framework \cite{Gillet2013, Delcorro2016, Wang2018}.

We now insert the Hamiltonians given in Equations~(\ref{eq:Hamiltonian1})-(\ref{eq:Hamiltonian3}) in the third-order perturbation rate, Equation~(\ref{eq:Pertubation}). As mentioned in the main text, for the Stokes processes involving one phonon, six permutations of $(\omega_\mathrm{in},-\omega_\mathrm{out},0)$ are used for $(\omega_1,\omega_2,\omega_3)$. Furthermore, the eigenstates of the unperturbed Hamiltonian $\ket{\Psi_{a}}$ (or $\ket{\Psi_{b}}$) can be written as $\ket{\psi_{e}} \otimes \ket{n_{\nu'}}$, where $\ket{\psi_{e}}$ and $\ket{n_{\nu'}}$ are the many-body electronic and phononic states, respectively (the index $e$ runs only over many-body electronic states). Since $\hat{H}_\mathrm{e\nu}$ and $\hat{H}_\mathrm{e\gamma}$ are, respectively, linear in and independent of the phononic operator, the phonon state $\ket{n_{\nu'}}$ contributes to the Stokes response only if $\ket{n_{\nu'}}$ is either $\ket{n_\nu}$ or $\ket{n_\nu\pm 1}$. Consequently, $\mel{\Psi_a}{\hat{H}_\mathrm{e\gamma}}{\Psi_i} = \mel{\psi_e}{\hat{H}_\mathrm{e\gamma}}{0} \delta_{\nu'\nu} \delta_{n_{\nu'}n_\nu}$ and $\mel{\Psi_a}{\hat{H}_\mathrm{e\nu}}{\Psi_i} = \mel{\psi_e}{\hat{H}_\mathrm{e\nu}}{0} \delta_{\nu'\nu} \delta_{n_{\nu'}(n_\nu\pm1)}$ ($\delta_{ij}$ denotes the Kronecker delta).
The total Raman intensity $I$ is obtained by summing over all final states, \ie phonon modes, and given by $I(\omega) = I_0 \sum_\nu (n_{\nu}+1)|P_{\nu}|^2 \delta(\omega-\omega_\nu)/\omega_\nu$, where $P_\nu$ is defined as
\begin{align}
\label{eq:Manybody}
P_{\nu} &\equiv \sum_{ed} \Bigg[ 
\frac{ \mel{0}{\hin}{\psi_e} \mel{\psi_e}{\gnu}{\psi_d} \mel{\psi_d}{\hout}{0} }{(\hbar\omega_\mathrm{in}-\ele{E}_e)(\hbar\omega_\mathrm{out}-\ele{E}_d)} 
+ \frac{ \mel{0}{\hin}{\psi_e} \mel{\psi_e}{\hout}{\psi_d} \mel{\psi_b}{\gnu}{0} }{(\hbar\omega_\mathrm{in}-\ele{E}_e)(\hbar\omega_\nu-\ele{E}_d)} \nonumber \\
&+ \frac{ \mel{0}{\hout}{\psi_e} \mel{\psi_e}{\gnu}{\psi_d} \mel{\psi_d}{\hin}{0} }{(-\hbar\omega_\mathrm{out}-\ele{E}_e)(-\hbar\omega_\mathrm{in}-\ele{E}_d)}
+ \frac{ \mel{0}{\hout}{\psi_e} \mel{\psi_e}{\hin}{\psi_d} \mel{\psi_d}{\gnu}{0} }{(-\hbar\omega_\mathrm{out}-\ele{E}_e)(\hbar\omega_\nu-\ele{E}_d)}  \\
&+ \frac{ \mel{0}{\gnu}{\psi_e} \mel{\psi_e}{\hin}{\psi_d} \mel{\psi_d}{\hout}{0} }{(-\hbar\omega_\nu-\ele{E}_e)(\hbar\omega_\mathrm{out}-\ele{E}_d)}
+\frac{ \mel{0}{\gnu}{\psi_e} \mel{\psi_e}{\hout}{\psi_d} \mel{\psi_d}{\hin}{0} }{(-\hbar\omega_\nu-\ele{E}_e)(-\hbar\omega_\mathrm{in}-\ele{E}_d)} \Bigg] \, . \nonumber
\end{align}	
Here, $\ele{E}_{e/d}$ denote the electronic energies (with respect to the electronic ground state), the summations over $e$ and $d$ include only electronic states, and $\hat{\va{P}}$ and $\gnu$ are many-body electronic operators given by $\hat{\va{P}} \equiv \sum_{nm\va{k}} \va{p}_{nm\va{k}} \hat{c}_{n\va{k}}^\dagger \hat{c}_{m\va{k}}$ and $\gnu \equiv \sum_{nm\va{k}} g_{nm\va{k}}^{\nu\va{0}} \hat{c}_{n\va{k}}^\dagger \hat{c}_{m\va{k}}$. Note that the momentum conservation implies that only phonons at $\va{q}=\va{0}$ contribute to the response here \cite{Saito2011}, \ie $\omega_\nu\equiv\omega_{\nu\va{0}}$. Since both $\hat{\va{P}}$ and $\gnu$ are bi-linear in the electronic operator, for a non-vanishing matrix elements, $\ket{\psi_{e/d}}$ must include singly-excited states, \ie terms in the form $\hat{c}_{c\va{k}}^\dagger\hat{c}_{v\va{k}}\ket{0}$ (indices $c$ and $v$ imply conduction and valence bands, respectively) \cite{Taghizadeh2018}. Excitonic effects can readily be introduced at this stage by incorporating the BSE solution \cite{Taghizadeh2018, Taghizadeh2019b}. However, we neglect the excitonic effects in the present work, and hence, each singly-excited state contributes individually to the response, \ie $\ket{\psi_{e/d}}= \hat{c}_{c\va{k}}^\dagger\hat{c}_{v\va{k}}\ket{0}$ with an energy of $\ele{E}_{e/d}=\varepsilon_{c\va{k}}-\varepsilon_{v\va{k}}$. At finite temperature, the expression for $\ket{\psi_{e/d}}$ should be taken as $\ket{\psi_{e/d}}=f_i(1-f_j)\hat{c}_{j\va{k}}^\dagger \hat{c}_{i\va{k}} \ket{0}$, where $f_i \equiv (1+\exp[(\varepsilon_{i\va{k}}-\mu)/k_BT])^{-1}$ is the Fermi--Dirac distribution with chemical potential $\mu$.

Rewriting $P_\nu$ in terms of the single-particle variables and polarization vectors leads to Equation~(\ref{eq:RamanFinal}) for the Raman intensity, where the Raman tensor component, $R_{\alpha\beta}^\nu$, reads 
\begin{align}
\label{eq:RamanTensor}
R_{\alpha\beta}^\nu \equiv \sum_{ijmn\va{k}} &\bigg[ \frac{p_{ij}^\alpha (g_{jm}^\nu \delta_{in}-g_{ni}^\nu \delta_{jm}) p_{mn}^\beta }{(\hbar\omega_\mathrm{in}-\varepsilon_{ji})(\hbar\omega_\mathrm{out}-\varepsilon_{mn})} + \frac{p_{ij}^\alpha (p_{jm}^\beta \delta_{in}-p_{ni}^\beta \delta_{jm}) g_{mn}^\nu }{(\hbar\omega_\mathrm{in}-\varepsilon_{ji})(\hbar\omega_\nu-\varepsilon_{mn})} \nonumber \\
&+\frac{p_{ij}^\beta (g_{jm}^\nu \delta_{in}-g_{ni}^\nu \delta_{jm}) p_{mn}^\alpha }{(-\hbar\omega_\mathrm{out}-\varepsilon_{ji})(-\hbar\omega_\mathrm{in}-\varepsilon_{mn})} 
+\frac{p_{ij}^\beta (p_{jm}^\alpha \delta_{in}-p_{ni}^\alpha \delta_{jm}) g_{mn}^\nu }{(-\hbar\omega_\mathrm{out}-\varepsilon_{ji})(\hbar\omega_\nu-\varepsilon_{mn})} \\
&+\frac{g_{ij}^\nu (p_{jm}^\alpha \delta_{in}-p_{ni}^\alpha \delta_{jm}) p_{mn}^\beta }{(-\hbar\omega_\nu-\varepsilon_{ji})(\hbar\omega_\mathrm{out}-\varepsilon_{mn})} + \frac{g_{ij}^\nu (p_{jm}^\beta \delta_{in}-p_{ni}^\beta \delta_{jm}) p_{mn}^\alpha }{(-\hbar\omega_\nu-\varepsilon_{ji})(-\hbar\omega_\mathrm{in}-\varepsilon_{mn})} \bigg] f_i(1-f_j)f_n(1-f_m)  \, . \nonumber
\end{align}
Here, $\varepsilon_{ij}\equiv \varepsilon_{i\va{k}}-\varepsilon_{j\va{k}}$, $p_{ij}^\alpha \equiv \mel{i\va{k}}{\hat{p}^\alpha}{j\va{k}}$, $g_{ij}^\nu \equiv \mel{i\va{k}}{\partial_{\nu\va{0}}V^\mathrm{KS}}{j\va{k}}$, and $(i,j,m,n)$/$\nu$ are the electron/phonon band index.   
The line-shape broadening is accounted for by adding a small phenomenological imaginary part, $i\eta$, to the photon frequencies $\omega_\mathrm{in/out} \rightarrow \omega_\mathrm{in/out}+i\eta$. We set the frequency broadening to $\eta=200$ meV in our calculations. 


\subsection*{First-principles Calculations}
All DFT calculations are performed with the projector augmented wave code, GPAW \cite{Mortensen2005, Enkovaara2010}, in combination with the atomic simulation environment (ASE) \cite{HjorthLarsen2017}. The Perdew-Burke-Ernzerhof (PBE) exchange-correlation functional is used \cite{Perdew1996} and the Kohn-Sham orbitals are expanded using the double zeta polarized (dzp) basis set \cite{Larsen2009}. Despite its fairly small size, the dzp basis set provides sufficiently accurate phonon modes. This has been tested by benchmarking the phonon frequencies obtained from this basis set against the results using the commonly-employed plane waves for 700+ monolayers (more than 7000 phonon modes). We confirm that for approximately 80\% of all phonons, the discrepancy between the two approaches is less than 5\%.	
Also, the choice of exchange-correlation functional may slightly influence the Raman spectra. For instance, it is known that the PBE functional tends to overestimate the lattice parameters and underestimate the phonon frequencies in crystals \cite{Zhang2018}, whereas the opposite occurs for the local-density approximation (LDA) functionals. Nonetheless, this choice only slightly influences our calculated Raman spectra, and PBE usually provides sufficiently accurate phonon frequencies in the range of theoretical and experimental uncertainties \cite{Skelton2015}. The monolayers are placed between two vacuum regions with thicknesses of 15 \AA. A convergence test of Raman spectra with respect to the wavevector density is performed for several materials, and a mesh with the density of 25 \AA$^{-1}$ for ground state calculations was chosen. The phonon modes are obtained using the standard approach based on calculating the dynamical matrices in the harmonic approximation \cite{Fritsch1999}. The dynamical matrix is evaluated using the small-displacement method \cite{Alf2009}, where the change of forces on a specific atom caused by varying the position of neighbouring atoms is computed. Since only the zone-centered ($\Gamma$-point) phonons are required, the phonon modes can be computed based on the crystal unit cell. A k-mesh with a density of 12 \AA$^{-1}$ is used for phonon calculations, and the forces are converged within 10$^{-6}$ eV\AA$^{-1}$. Since the wavefunctions and Kohn-Sham potentials in GPAW are evaluated on a real-space grid \cite{Mortensen2005}, a convergence test with respect to this grid spacing is performed and a real-space grid of 0.2 {\AA} is chosen for calculations. The electron-phonon matrix elements are then obtained within the adiabatic approximation using a finite difference technique for evaluating the derivative of the Kohn-Sham potential \cite{Kaasbjerg2012}. Similarly, the momentum matrix elements are calculated using the finite difference technique and the correction terms due to projector augmented waves \cite{Blochl1994} are added. The width of the Fermi-Dirac occupations is set to $k_BT=50$ meV for faster convergence of the DFT results. For generating the Raman spectra, a Gaussian [$G(\omega)=(\sigma\sqrt{2\pi})^{-1}\exp{(-\omega^2/2\sigma^2)}$] with a variance $\sigma=3$ cm$^{-1}$ is used to replace the Dirac delta function, which accounts for the inhomogeneous broadening of phonon modes. The temperature of the Bose-Einstein distributions is set to 300 K for all calculations except for the results in top panel of Fig.~\ref{fig:Distances}, where a temperature of 77 K is used. The calculations are submitted, managed, and received using the simple MyQueue workflow tool \cite{Mortensen2020}, which is a Python front-end to job scheduler.


\subsection*{Experimental Raman Spectra}
The experimental Raman spectra are extracted from the figures in the corresponding references using a common plot digitizer. To remove the noise in the experimental data, they are filtered using a Savitzky-Golay filter \cite{Savitzky1964} of order three with a filter window length of eleven. For a fair comparison with our theoretical spectra in Fig.~\ref{fig:ExpCompare}, we have convolved the experimental spectra with a Gaussian function with variance of 10 cm$^{-1}$ to reduce the effect of possible but unimportant small frequency shifts between the experimental and theoretical spectra. Furthermore, the Raman moments have been calculated over a frequency range where the main Raman peaks appear, from 350 to 450 cm$^{-1}$ for MoS$_2$ and from 75 to 260 cm$^{-1}$ for WTe$_2$. 
For calculating the Euclidean distance, both the experimental and theoretical spectra are convolved with a Gaussian function with variance of 6 cm$^{-1}$. 


\section*{Data Availability}
All calculated Raman spectra are freely available online through the \href{https://cmrdb.fysik.dtu.dk/c2db/}{C2DB website}. Other data is available from the corresponding author upon reasonable request. 


\section*{Code Availability}
GPAW is an open-source DFT Python code based on the projector-augmented wave method and the ASE, which is available at the \href{https://wiki.fysik.dtu.dk/gpaw/}{GPAW website}. The Raman code used for generating Raman spectra in this work will be available in future releases of the code.




\section*{Acknowledgments}
The authors thank M. N. Gjerding for helpful discussions throughout the project. This work was supported by the Center for Nanostructured Graphene (CNG) under the Danish National Research Foundation (project DNRF103). K. S. T. acknowledges support from the European Research Council (ERC) under the European Union’s Horizon 2020 research and innovation program (Grant No. 773122, LIMA).


\section*{Author Contributions}
U.L. wrote the initial computer code. A.T. developed the formalism, modified the code and carried out the calculations. All authors discussed the results. A.T., T.G.P. and K.S.T. wrote the manuscript. T.G.P. and K.S.T. supervised the project. 


\section*{Additional information}
\noindent\textbf{Supplementary Information:} accompanies this paper.

\noindent\textbf{Competing interests:} The authors declare no competing interests.


\bibliographystyle{ieeetr}
\bibliography{main}

\begin{thebibliography}{10}

\bibitem{Novoselov2004}
K.~S. Novoselov, A.~K. Geim, S.~V. Morozov, D.~Jiang, Y.~Zhang, S.~V. Dubonos,
  I.~V. Grigorieva, and A.~A. Firsov, ``Electric field effect in atomically
  thin carbon films,'' {\em Science}, vol.~306, no.~5696, pp.~666--669, 2004.

\bibitem{Anasori2017}
B.~Anasori, M.~R. Lukatskaya, and Y.~Gogotsi, ``{2D} metal carbides and
  nitrides ({MXenes}) for energy storage,'' {\em Nat. Rev. Mater.}, vol.~2,
  no.~2, p.~16098, 2017.

\bibitem{Xi2015}
X.~Xi, L.~Zhao, Z.~Wang, H.~Berger, L.~Forr{\'{o}}, J.~Shan, and K.~F. Mak,
  ``Strongly enhanced charge-density-wave order in monolayer {NbSe}$_2$,'' {\em
  Nat. Nanotechnol.}, vol.~10, no.~9, pp.~765--769, 2015.

\bibitem{Liu2014}
H.~Liu, A.~T. Neal, Z.~Zhu, Z.~Luo, X.~Xu, D.~Tom{\'{a}}nek, and P.~D. Ye,
  ``Phosphorene: An unexplored {2D} semiconductor with a high hole mobility,''
  {\em {ACS} Nano}, vol.~8, no.~4, pp.~4033--4041, 2014.

\bibitem{Mak2010}
K.~F. Mak, C.~Lee, J.~Hone, J.~Shan, and T.~F. Heinz, ``Atomically thin
  {MoS}$_2$: A new direct-gap semiconductor,'' {\em Phys. Rev. Lett.},
  vol.~105, p.~136805, 2010.

\bibitem{Wang2012}
Q.~H. Wang, K.~Kalantar-Zadeh, A.~Kis, J.~N. Coleman, and M.~S. Strano,
  ``Electronics and optoelectronics of two-dimensional transition metal
  dichalcogenides,'' {\em Nat. Nanotechnol.}, vol.~7, no.~11, pp.~699--712,
  2012.

\bibitem{Ci2010}
L.~Ci, L.~Song, C.~Jin, D.~Jariwala, D.~Wu, Y.~Li, A.~Srivastava, Z.~F. Wang,
  K.~Storr, L.~Balicas, F.~Liu, and P.~M. Ajayan, ``Atomic layers of hybridized
  boron nitride and graphene domains,'' {\em Nat. Mater.}, vol.~9, no.~5,
  pp.~430--435, 2010.

\bibitem{Huang2017}
B.~Huang, G.~Clark, E.~Navarro-Moratalla, D.~R. Klein, R.~Cheng, K.~L. Seyler,
  D.~Zhong, E.~Schmidgall, M.~A. McGuire, D.~H. Cobden, W.~Yao, D.~Xiao,
  P.~Jarillo-Herrero, and X.~Xu, ``Layer-dependent ferromagnetism in a {van der
  Waals} crystal down to the monolayer limit,'' {\em Nature}, vol.~546,
  no.~7657, pp.~270--273, 2017.

\bibitem{Wang2017}
H.~Wang, X.~Huang, J.~Lin, J.~Cui, Y.~Chen, C.~Zhu, F.~Liu, Q.~Zeng, J.~Zhou,
  P.~Yu, X.~Wang, H.~He, S.~H. Tsang, W.~Gao, K.~Suenaga, F.~Ma, C.~Yang,
  L.~Lu, T.~Yu, E.~H.~T. Teo, G.~Liu, and Z.~Liu, ``High-quality monolayer
  superconductor {NbSe}$_2$ grown by chemical vapour deposition,'' {\em Nat.
  Commun.}, vol.~8, no.~1, p.~394, 2017.

\bibitem{Cao2018}
Y.~Cao, V.~Fatemi, S.~Fang, K.~Watanabe, T.~Taniguchi, E.~Kaxiras, and
  P.~Jarillo-Herrero, ``Unconventional superconductivity in magic-angle
  graphene superlattices,'' {\em Nature}, vol.~556, no.~7699, pp.~43--50, 2018.

\bibitem{Haastrup2018}
S.~Haastrup, M.~Strange, M.~Pandey, T.~Deilmann, P.~S. Schmidt, N.~F. Hinsche,
  M.~N. Gjerding, D.~Torelli, P.~M. Larsen, A.~C. Riis-Jensen, J.~Gath, K.~W.
  Jacobsen, J.~J. Mortensen, T.~Olsen, and K.~S. Thygesen, ``{The Computational
  2D Materials Database}: high-throughput modeling and discovery of atomically
  thin crystals,'' {\em 2D Mater.}, vol.~5, no.~4, p.~042002, 2018.

\bibitem{Bonaccorso2010}
F.~Bonaccorso, Z.~Sun, T.~Hasan, and A.~C. Ferrari, ``Graphene photonics and
  optoelectronics,'' {\em Nat. Photon.}, vol.~4, no.~9, pp.~611--622, 2010.

\bibitem{Gjerding2017}
M.~N. Gjerding, R.~Petersen, T.~G. Pedersen, N.~A. Mortensen, and K.~S.
  Thygesen, ``Layered van der {Waals} crystals with hyperbolic light
  dispersion,'' {\em Nat. Commun.}, vol.~8, no.~1, p.~320, 2017.

\bibitem{Ferrari2007}
A.~C. Ferrari, ``Raman spectroscopy of graphene and graphite: Disorder,
  electron-phonon coupling, doping and nonadiabatic effects,'' {\em Solid State
  Commun.}, vol.~143, no.~1-2, pp.~47--57, 2007.

\bibitem{Ferrari2013}
A.~C. Ferrari and D.~M. Basko, ``Raman spectroscopy as a versatile tool for
  studying the properties of graphene,'' {\em Nat. Nanotechnol.}, vol.~8,
  no.~4, pp.~235--246, 2013.

\bibitem{Neumann2015}
C.~Neumann, S.~Reichardt, P.~Venezuela, M.~Dr\"{o}geler, L.~Banszerus,
  M.~Schmitz, K.~Watanabe, T.~Taniguchi, F.~Mauri, B.~Beschoten, S.~V. Rotkin,
  and C.~Stampfer, ``Raman spectroscopy as probe of nanometre-scale strain
  variations in graphene,'' {\em Nat. Commun.}, vol.~6, no.~1, p.~8429, 2015.

\bibitem{Li2017}
X.-L. Li, W.-P. Han, J.-B. Wu, X.-F. Qiao, J.~Zhang, and P.-H. Tan,
  ``Layer-number dependent optical properties of {2D} materials and their
  application for thickness determination,'' {\em Adv. Funct. Mater.}, vol.~27,
  no.~19, p.~1604468, 2017.

\bibitem{Long2002}
D.~A. Long, {\em The Raman Effect: A Unified Treatment of the Theory of {Raman}
  Scattering by Molecules}.
\newblock John Wiley {\&} Sons Ltd, Chichester, 2002.

\bibitem{Saito2011}
R.~Saito, M.~Hofmann, G.~Dresselhaus, A.~Jorio, and M.~S. Dresselhaus, ``Raman
  spectroscopy of graphene and carbon nanotubes,'' {\em Adv. Phys.}, vol.~60,
  no.~3, pp.~413--550, 2011.

\bibitem{Wilson1980}
C.~P. Wilson~E.B., Decius~J.C., ed., {\em Molecular vibrations. The theory of
  infrared and {Raman} vibrational spectra}.
\newblock Dover Publications, New York, 1980.

\bibitem{Umari2003}
P.~Umari and A.~Pasquarello, ``First-principles analysis of the {Raman}
  spectrum of vitreous silica: comparison with the vibrational density of
  states,'' {\em J. Phys. Condens. Matter}, vol.~15, no.~16, pp.~S1547--S1552,
  2003.

\bibitem{Liang2019}
Q.~Liang, S.~Dwaraknath, and K.~A. Persson, ``High-throughput computation and
  evaluation of {Raman} spectra,'' {\em Sci. Data}, vol.~6, no.~1, p.~135,
  2019.

\bibitem{Cai2017}
Q.~Cai, D.~Scullion, A.~Falin, K.~Watanabe, T.~Taniguchi, Y.~Chen, E.~J.~G.
  Santos, and L.~H. Li, ``Raman signature and phonon dispersion of atomically
  thin boron nitride,'' {\em Nanoscale}, vol.~9, no.~9, pp.~3059--3067, 2017.

\bibitem{Jiang2016}
Y.~C. Jiang, J.~Gao, and L.~Wang, ``Raman fingerprint for semi-metal {WTe}$_2$
  evolving from bulk to monolayer,'' {\em Sci. Rep.}, vol.~6, no.~1, p.~19624,
  2016.

\bibitem{Dewandre2019}
A.~Dewandre, M.~J. Verstraete, N.~Grobert, and Z.~Zanolli, ``Spectroscopic
  properties of few-layer tin chalcogenides,'' {\em J. Phys.: Mater.}, vol.~2,
  no.~4, p.~044005, 2019.

\bibitem{Liang2014}
L.~Liang and V.~Meunier, ``First-principles {Raman} spectra of {MoS}$_2$,
  {WS}$_2$ and their heterostructures,'' {\em Nanoscale}, vol.~6, no.~10,
  p.~5394, 2014.

\bibitem{Larsen2009}
A.~H. Larsen, M.~Vanin, J.~J. Mortensen, K.~S. Thygesen, and K.~W. Jacobsen,
  ``Localized atomic basis set in the projector augmented wave method,'' {\em
  Phys. Rev. B}, vol.~80, p.~195112, 2009.

\bibitem{C2DB}
``The {C2DB} website.''
\newblock \url{https://cmrdb.fysik.dtu.dk/c2db}.

\bibitem{Butler2018}
K.~T. Butler, D.~W. Davies, H.~Cartwright, O.~Isayev, and A.~Walsh, ``Machine
  learning for molecular and materials science,'' {\em Nature}, vol.~559,
  no.~7715, pp.~547--555, 2018.

\bibitem{Schmidt2019}
J.~Schmidt, M.~R.~G. Marques, S.~Botti, and M.~A.~L. Marques, ``Recent advances
  and applications of machine learning in solid-state materials science,'' {\em
  Npj Comput. Mater.}, vol.~5, no.~1, p.~83, 2019.

\bibitem{Albrecht1961}
A.~C. Albrecht, ``On the theory of {Raman} intensities,'' {\em J. Chem. Phys.},
  vol.~34, no.~5, pp.~1476--1484, 1961.

\bibitem{Lee1979}
S.-Y. Lee and E.~J. Heller, ``Time-dependent theory of {Raman} scattering,''
  {\em J. Chem. Phys.}, vol.~71, no.~12, p.~4777, 1979.

\bibitem{Ganguly1967}
A.~K. Ganguly and J.~L. Birman, ``Theory of lattice {Raman} scattering in
  insulators,'' {\em Phys. Rev.}, vol.~162, pp.~806--816, 1967.

\bibitem{Yu2010}
P.~Y. Yu and M.~Cardona, {\em Fundamentals of semiconductors: physics and
  materials properties}.
\newblock Springer, Berlin, 2010.

\bibitem{Grosso2000}
G.~Grosso and G.~P. Parravicini, {\em Solid State Physics}.
\newblock Academic Press, San Diego, California, 2000.

\bibitem{Zhang2017}
J.~Zhang, S.~Jia, I.~Kholmanov, L.~Dong, D.~Er, W.~Chen, H.~Guo, Z.~Jin, V.~B.
  Shenoy, L.~Shi, and J.~Lou, ``Janus monolayer transition-metal
  dichalcogenides,'' {\em {ACS} Nano}, vol.~11, no.~8, pp.~8192--8198, 2017.

\bibitem{Loudon1964}
R.~Loudon, ``The {Raman} effect in crystals,'' {\em Adv. Phys.}, vol.~13,
  no.~52, pp.~423--482, 1964.

\bibitem{Ribeiro2014}
J.~Ribeiro-Soares, R.~M. Almeida, E.~B. Barros, P.~T. Araujo, M.~S.
  Dresselhaus, L.~G. Can\ifmmode~\mbox{\c{c}}\else \c{c}\fi{}ado, and A.~Jorio,
  ``Group theory analysis of phonons in two-dimensional transition metal
  dichalcogenides,'' {\em Phys. Rev. B}, vol.~90, p.~115438, 2014.

\bibitem{Wu2018}
J.-B. Wu, M.-L. Lin, X.~Cong, H.-N. Liu, and P.-H. Tan, ``Raman spectroscopy of
  graphene-based materials and its applications in related devices,'' {\em
  Chem. Soc. Rev}, vol.~47, no.~5, pp.~1822--1873, 2018.

\bibitem{Xia2014}
F.~Xia, H.~Wang, and Y.~Jia, ``Rediscovering black phosphorus as an anisotropic
  layered material for optoelectronics and electronics,'' {\em Nat. Commun.},
  vol.~5, no.~1, p.~4458, 2014.

\bibitem{Ruppert2014}
C.~Ruppert, O.~B. Aslan, and T.~F. Heinz, ``Optical properties and band gap of
  single- and few-layer {MoTe}$_2$ crystals,'' {\em Nano Lett.}, vol.~14,
  no.~11, pp.~6231--6236, 2014.

\bibitem{Liu2018}
H.-L. Liu, T.~Yang, Y.~Tatsumi, Y.~Zhang, B.~Dong, H.~Guo, Z.~Zhang,
  Y.~Kumamoto, M.-Y. Li, L.-J. Li, R.~Saito, and S.~Kawata, ``Deep-ultraviolet
  {Raman} scattering spectroscopy of monolayer {WS}$_2$,'' {\em Sci. Rep.},
  vol.~8, no.~1, p.~11398, 2018.

\bibitem{Yang2017}
M.~Yang, X.~Cheng, Y.~Li, Y.~Ren, M.~Liu, and Z.~Qi, ``Anharmonicity of
  monolayer {MoS}$_2$, {MoSe}$_2$, and {WSe}$_2$: A {Raman} study under high
  pressure and elevated temperature,'' {\em Appl. Phys. Lett.}, vol.~110,
  no.~9, p.~093108, 2017.

\bibitem{Zhao2016b}
Y.~Zhao, J.~Qiao, P.~Yu, Z.~Hu, Z.~Lin, S.~P. Lau, Z.~Liu, W.~Ji, and Y.~Chai,
  ``Extraordinarily strong interlayer interaction in {2D} layered {PtS}$_2$,''
  {\em Adv. Mater.}, vol.~28, no.~12, pp.~2399--2407, 2016.

\bibitem{Yan2017}
M.~Yan, E.~Wang, X.~Zhou, G.~Zhang, H.~Zhang, K.~Zhang, W.~Yao, N.~Lu, S.~Yang,
  S.~Wu, T.~Yoshikawa, K.~Miyamoto, T.~Okuda, Y.~Wu, P.~Yu, W.~Duan, and
  S.~Zhou, ``High quality atomically thin {PtSe}$_2$ films grown by molecular
  beam epitaxy,'' {\em 2D Mater.}, vol.~4, no.~4, p.~045015, 2017.

\bibitem{Gupta2014}
U.~Gupta, B.~S. Naidu, U.~Maitra, A.~Singh, S.~N. Shirodkar, U.~V. Waghmare,
  and C.~N.~R. Rao, ``Characterization of few-layer {1T}-{MoSe}$_2$ and its
  superior performance in the visible-light induced hydrogen evolution
  reaction,'' {\em {APL} Mater.}, vol.~2, no.~9, p.~092802, 2014.

\bibitem{Chen2017}
K.~Chen, Z.~Chen, X.~Wan, Z.~Zheng, F.~Xie, W.~Chen, X.~Gui, H.~Chen, W.~Xie,
  and J.~Xu, ``A simple method for synthesis of high-quality millimeter-scale
  {{1T}$^\prime$} transition-metal telluride and near-field nanooptical
  properties,'' {\em Adv. Mater.}, vol.~29, no.~38, p.~1700704, 2017.

\bibitem{Cao2017}
Y.~Cao, N.~Sheremetyeva, L.~Liang, H.~Yuan, T.~Zhong, V.~Meunier, and M.~Pan,
  ``Anomalous vibrational modes in few layer {WTe}$_2$ revealed by polarized
  {Raman} scattering and first-principles calculations,'' {\em 2D Mater.},
  vol.~4, no.~3, p.~035024, 2017.

\bibitem{Yu2020}
J.~Yu, X.~Kuang, Y.~Gao, Y.~Wang, K.~Chen, Z.~Ding, J.~Liu, C.~Cong, J.~He,
  Z.~Liu, and Y.~Liu, ``Direct observation of the linear dichroism transition
  in two-dimensional palladium diselenide,'' {\em Nano Lett.}, vol.~20, no.~2,
  pp.~1172--1182, 2020.

\bibitem{Carvalho2015}
B.~R. Carvalho, L.~M. Malard, J.~M. Alves, C.~Fantini, and M.~A. Pimenta,
  ``Symmetry-dependent exciton-phonon coupling in {2D} and bulk {MoS}$_2$
  observed by resonance {Raman} scattering,'' {\em Phys. Rev. Lett.}, vol.~114,
  p.~136403, 2015.

\bibitem{Scheuschner2015}
N.~Scheuschner, R.~Gillen, M.~Staiger, and J.~Maultzsch, ``Interlayer resonant
  {Raman} modes in few-layer {MoS}$_2$,'' {\em Phys. Rev. B}, vol.~91,
  p.~235409, 2015.

\bibitem{Lee2015}
J.-U. Lee, K.~Kim, and H.~Cheong, ``Resonant {Raman} and photoluminescence
  spectra of suspended molybdenum disulfide,'' {\em 2D Mater.}, vol.~2, no.~4,
  p.~044003, 2015.

\bibitem{Delcorro2016}
E.~del Corro, A.~Botello-M{\'{e}}ndez, Y.~Gillet, A.~L. Elias, H.~Terrones,
  S.~Feng, C.~Fantini, D.~Rhodes, N.~Pradhan, L.~Balicas, X.~Gonze, J.-C.
  Charlier, M.~Terrones, and M.~A. Pimenta, ``Atypical exciton--phonon
  interactions in {WS}$_2$ and {WSe}$_2$ monolayers revealed by resonance
  {Raman} spectroscopy,'' {\em Nano Lett.}, vol.~16, no.~4, pp.~2363--2368,
  2016.

\bibitem{Gillet2013}
Y.~Gillet, M.~Giantomassi, and X.~Gonze, ``First-principles study of excitonic
  effects in {Raman} intensities,'' {\em Phys. Rev. B}, vol.~88, p.~094305,
  2013.

\bibitem{Wang2018}
Y.~Wang, B.~R. Carvalho, and V.~H. Crespi, ``Strong exciton regulation of
  {Raman} scattering in monolayer {MoS}$_2$,'' {\em Phys. Rev. B}, vol.~98,
  p.~161405, 2018.

\bibitem{Taghizadeh2019}
A.~Taghizadeh and T.~G. Pedersen, ``Nonlinear optical selection rules of
  excitons in monolayer transition metal dichalcogenides,'' {\em Phys. Rev. B},
  vol.~99, p.~235433, 2019.

\bibitem{Taghizadeh2019b}
A.~Taghizadeh and T.~G. Pedersen, ``Nonlinear excitonic spin {Hall} effect in
  monolayer transition metal dichalcogenides,'' {\em 2D Mater.}, vol.~7, no.~1,
  p.~015003, 2019.

\bibitem{Li2012}
H.~Li, Q.~Zhang, C.~C.~R. Yap, B.~K. Tay, T.~H.~T. Edwin, A.~Olivier, and
  D.~Baillargeat, ``From bulk to monolayer {MoS}$_2$: Evolution of {Raman}
  scattering,'' {\em Adv. Funct. Mater.}, vol.~22, no.~7, pp.~1385--1390, 2012.

\bibitem{Lanzillo2013}
N.~A. Lanzillo, A.~G. Birdwell, M.~Amani, F.~J. Crowne, P.~B. Shah, S.~Najmaei,
  Z.~Liu, P.~M. Ajayan, J.~Lou, M.~Dubey, S.~K. Nayak, and T.~P. O'Regan,
  ``Temperature-dependent phonon shifts in monolayer {MoS}$_2$,'' {\em Appl.
  Phys. Lett.}, vol.~103, no.~9, p.~093102, 2013.

\bibitem{Tongay2013}
S.~Tongay, J.~Suh, C.~Ataca, W.~Fan, A.~Luce, J.~S. Kang, J.~Liu, C.~Ko,
  R.~Raghunathanan, J.~Zhou, F.~Ogletree, J.~Li, J.~C. Grossman, and J.~Wu,
  ``Defects activated photoluminescence in two-dimensional semiconductors:
  interplay between bound, charged and free excitons,'' {\em Sci. Rep}, vol.~3,
  no.~1, p.~2657, 2013.

\bibitem{Zhou2016}
J.~Zhou, F.~Liu, J.~Lin, X.~Huang, J.~Xia, B.~Zhang, Q.~Zeng, H.~Wang, C.~Zhu,
  L.~Niu, X.~Wang, W.~Fu, P.~Yu, T.-R. Chang, C.-H. Hsu, D.~Wu, H.-T. Jeng,
  Y.~Huang, H.~Lin, Z.~Shen, C.~Yang, L.~Lu, K.~Suenaga, W.~Zhou, S.~T.
  Pantelides, G.~Liu, and Z.~Liu, ``Large-area and high-quality {2D} transition
  metal telluride,'' {\em Adv. Mater.}, vol.~29, no.~3, p.~1603471, 2016.

\bibitem{Ghosh2019}
K.~Ghosh, A.~Stuke, M.~Todorovi{\'{c}}, P.~B. J{\o}rgensen, M.~N. Schmidt,
  A.~Vehtari, and P.~Rinke, ``Deep learning spectroscopy: Neural networks for
  molecular excitation spectra,'' {\em Adv. Sci.}, vol.~6, no.~9, p.~1801367,
  2019.

\bibitem{Cudazzo2016}
P.~Cudazzo, L.~Sponza, C.~Giorgetti, L.~Reining, F.~Sottile, and M.~Gatti,
  ``Exciton band structure in two-dimensional materials,'' {\em Phys. Rev.
  Lett.}, vol.~116, p.~066803, 2016.

\bibitem{Taghizadeh2018}
A.~Taghizadeh and T.~G. Pedersen, ``Gauge invariance of excitonic linear and
  nonlinear optical response,'' {\em Phys. Rev. B}, vol.~97, p.~205432, 2018.

\bibitem{Liang2017}
L.~Liang, J.~Zhang, B.~G. Sumpter, Q.-H. Tan, P.-H. Tan, and V.~Meunier,
  ``Low-frequency shear and layer-breathing modes in {Raman} scattering of
  two-dimensional materials,'' {\em {ACS} Nano}, vol.~11, no.~12,
  pp.~11777--11802, 2017.

\bibitem{Malard2009}
L.~M. Malard, M.~H.~D. Guimar\~aes, D.~L. Mafra, M.~S.~C. Mazzoni, and
  A.~Jorio, ``Group-theory analysis of electrons and phonons in $n$-layer
  graphene systems,'' {\em Phys. Rev. B}, vol.~79, p.~125426, 2009.

\bibitem{Ribeiro2015}
J.~Ribeiro-Soares, R.~M. Almeida, L.~G. Can\ifmmode~\mbox{\c{c}}\else
  \c{c}\fi{}ado, M.~S. Dresselhaus, and A.~Jorio, ``Group theory for structural
  analysis and lattice vibrations in phosphorene systems,'' {\em Phys. Rev. B},
  vol.~91, p.~205421, 2015.

\bibitem{Taghizadeh2017}
A.~Taghizadeh, F.~Hipolito, and T.~G. Pedersen, ``Linear and nonlinear optical
  response of crystals using length and velocity gauges: Effect of basis
  truncation,'' {\em Phys. Rev. B}, vol.~96, p.~195413, 2017.

\bibitem{Reichardt2019}
S.~Reichardt and L.~Wirtz, ``Theory of resonant {Raman} scattering: Towards a
  comprehensive \textit{ab initio} description,'' {\em Phys. Rev. B}, vol.~99,
  p.~174312, 2019.

\bibitem{Giustino2017}
F.~Giustino, ``Electron-phonon interactions from first principles,'' {\em Rev.
  Mod. Phys.}, vol.~89, p.~015003, 2017.

\bibitem{Mortensen2005}
J.~J. Mortensen, L.~B. Hansen, and K.~W. Jacobsen, ``Real-space grid
  implementation of the projector augmented wave method,'' {\em Phys. Rev. B},
  vol.~71, p.~035109, 2005.

\bibitem{Enkovaara2010}
J.~Enkovaara, C.~Rostgaard, J.~J. Mortensen, J.~Chen, M.~Du{\l}ak, L.~Ferrighi,
  J.~Gavnholt, C.~Glinsvad, V.~Haikola, H.~A. Hansen, H.~H. Kristoffersen,
  M.~Kuisma, A.~H. Larsen, L.~Lehtovaara, M.~Ljungberg, O.~Lopez-Acevedo, P.~G.
  Moses, J.~Ojanen, T.~Olsen, V.~Petzold, N.~A. Romero, J.~Stausholm-M{\o}ller,
  M.~Strange, G.~A. Tritsaris, M.~Vanin, M.~Walter, B.~Hammer,
  H.~H{\"{a}}kkinen, G.~K.~H. Madsen, R.~M. Nieminen, J.~K. N{\o}rskov,
  M.~Puska, T.~T. Rantala, J.~Schi{\o}tz, K.~S. Thygesen, and K.~W. Jacobsen,
  ``Electronic structure calculations with {GPAW}: a real-space implementation
  of the projector augmented-wave method,'' {\em J. Condens. Matter Phys.},
  vol.~22, no.~25, p.~253202, 2010.

\bibitem{HjorthLarsen2017}
A.~H. Larsen, J.~J. Mortensen, J.~Blomqvist, I.~E. Castelli, R.~Christensen,
  M.~Du{\l}ak, J.~Friis, M.~N. Groves, B.~Hammer, C.~Hargus, E.~D. Hermes,
  P.~C. Jennings, P.~B. Jensen, J.~Kermode, J.~R. Kitchin, E.~L. Kolsbjerg,
  J.~Kubal, K.~Kaasbjerg, S.~Lysgaard, J.~B. Maronsson, T.~Maxson, T.~Olsen,
  L.~Pastewka, A.~Peterson, C.~Rostgaard, J.~Schi{\o}tz, O.~Sch{\"{u}}tt,
  M.~Strange, K.~S. Thygesen, T.~Vegge, L.~Vilhelmsen, M.~Walter, Z.~Zeng, and
  K.~W. Jacobsen, ``The atomic simulation environment: a {Python} library for
  working with atoms,'' {\em J. Phys. Condens. Matter}, vol.~29, no.~27,
  p.~273002, 2017.

\bibitem{Perdew1996}
J.~P. Perdew, K.~Burke, and M.~Ernzerhof, ``Generalized gradient approximation
  made simple,'' {\em Phys. Rev. Lett.}, vol.~77, pp.~3865--3868, 1996.

\bibitem{Zhang2018}
G.-X. Zhang, A.~M. Reilly, A.~Tkatchenko, and M.~Scheffler, ``Performance of
  various density-functional approximations for cohesive properties of 64 bulk
  solids,'' {\em New J. Phys.}, vol.~20, no.~6, p.~063020, 2018.

\bibitem{Skelton2015}
J.~M. Skelton, D.~Tiana, S.~C. Parker, A.~Togo, I.~Tanaka, and A.~Walsh,
  ``Influence of the exchange-correlation functional on the quasi-harmonic
  lattice dynamics of {II}-{VI} semiconductors,'' {\em J. Chem. Phys.},
  vol.~143, no.~6, p.~064710, 2015.

\bibitem{Fritsch1999}
J.~Fritsch, ``Density functional calculation of semiconductor surface
  phonons,'' {\em Phys. Rep.}, vol.~309, no.~4-6, pp.~209--331, 1999.

\bibitem{Alf2009}
D.~Alf{\`{e}}, ``{PHON}: A program to calculate phonons using the small
  displacement method,'' {\em Comput. Phys. Commun.}, vol.~180, no.~12,
  pp.~2622--2633, 2009.

\bibitem{Kaasbjerg2012}
K.~Kaasbjerg, K.~S. Thygesen, and K.~W. Jacobsen, ``Phonon-limited mobility in
  $n$-type single-layer {MoS}$_2$ from first principles,'' {\em Phys. Rev. B},
  vol.~85, p.~115317, 2012.

\bibitem{Blochl1994}
P.~E. Bl{\"{o}}chl, ``Projector augmented-wave method,'' {\em Phys. Rev. B},
  vol.~50, pp.~17953--17979, 1994.

\bibitem{Mortensen2020}
J.~Mortensen, M.~Gjerding, and K.~Thygesen, ``{MyQueue}: Task and workflow
  scheduling system,'' {\em J. Open Source Softw.}, vol.~5, no.~45, p.~1844,
  2020.

\bibitem{Savitzky1964}
A.~Savitzky and M.~J.~E. Golay, ``Smoothing and differentiation of data by
  simplified least squares procedures.,'' {\em Anal. Chem.}, vol.~36, no.~8,
  pp.~1627--1639, 1964.

\end{thebibliography}

\end{document}